\documentclass[12pt]{article}
\usepackage{amsmath}
\usepackage{graphicx,psfrag,epsf}
\usepackage{enumerate}
\usepackage{natbib}
\usepackage{textcomp}
\usepackage[hyphens]{url} 
\usepackage{hyperref}

\newcommand{\blind}{0}

\usepackage{longtable,booktabs,array}
\usepackage{calc} 
\usepackage{etoolbox}
\makeatletter
\patchcmd\longtable{\par}{\if@noskipsec\mbox{}\fi\par}{}{}
\makeatother
\IfFileExists{footnotehyper.sty}{\usepackage{footnotehyper}}{\usepackage{footnote}}
\makesavenoteenv{longtable}

\usepackage{graphics}
\usepackage{float}
\usepackage{bm}
\usepackage{bbm}
\usepackage{amsmath}
\usepackage{amsfonts}
\usepackage{multirow}
\usepackage{booktabs}
\usepackage{longtable}
\usepackage{array}
\usepackage{wrapfig}
\usepackage{colortbl}
\usepackage{pdflscape}
\usepackage{tabu}
\usepackage{threeparttable}
\usepackage{threeparttablex}
\usepackage[normalem]{ulem}
\usepackage{makecell}
\usepackage{xcolor}

\usepackage{color}
\usepackage{fancyvrb}

\DefineVerbatimEnvironment{Highlighting}{Verbatim}{commandchars=\\\{\}}
\usepackage{framed}
\definecolor{shadecolor}{RGB}{248,248,248}
\newenvironment{Shaded}{\begin{snugshade}}{\end{snugshade}}

\newcommand{\AttributeTok}[1]{\textcolor[rgb]{0.77,0.63,0.00}{#1}}

\newcommand{\CommentTok}[1]{\textcolor[rgb]{0.56,0.35,0.01}{\textit{#1}}}

\newcommand{\ConstantTok}[1]{\textcolor[rgb]{0.00,0.00,0.00}{#1}}
\newcommand{\ControlFlowTok}[1]{\textcolor[rgb]{0.13,0.29,0.53}{\textbf{#1}}}
\newcommand{\DataTypeTok}[1]{\textcolor[rgb]{0.13,0.29,0.53}{#1}}
\newcommand{\DecValTok}[1]{\textcolor[rgb]{0.00,0.00,0.81}{#1}}
\newcommand{\DocumentationTok}[1]{\textcolor[rgb]{0.56,0.35,0.01}{\textbf{\textit{#1}}}}

\newcommand{\FloatTok}[1]{\textcolor[rgb]{0.00,0.00,0.81}{#1}}
\newcommand{\FunctionTok}[1]{\textcolor[rgb]{0.00,0.00,0.00}{#1}}

\newcommand{\KeywordTok}[1]{\textcolor[rgb]{0.13,0.29,0.53}{\textbf{#1}}}
\newcommand{\NormalTok}[1]{#1}

\newcommand{\OtherTok}[1]{\textcolor[rgb]{0.56,0.35,0.01}{#1}}

\newcommand{\SpecialCharTok}[1]{\textcolor[rgb]{0.00,0.00,0.00}{#1}}

\newcommand{\StringTok}[1]{\textcolor[rgb]{0.31,0.60,0.02}{#1}}

\begin{document}

\def\spacingset#1{\renewcommand{\baselinestretch}%
{#1}\small\normalsize} \spacingset{1}


\if0\blind
{
  \title{\bf Introducing Variational Inference in Statistics and Data Science Curriculum}

  \author{
        Vojtech Kejzlar \\
    Department of Mathematics and Statistics, Skidmore College\\
     and \\     Jingchen Hu \\
    Department of Mathematics and Statistics, Vassar College\\
      }
  \maketitle
} \fi

\if1\blind
{
  \bigskip
  \bigskip
  \bigskip
  \begin{center}
    {\LARGE\bf Introducing Variational Inference in Statistics and Data Science Curriculum}
  \end{center}
  \medskip
} \fi

\bigskip
\begin{abstract}
Probabilistic models such as logistic regression, Bayesian classification, neural networks, and models for natural language processing, are increasingly more present in both undergraduate and graduate statistics and data science curricula due to their wide range of applications. In this paper, we present a one-week course module for studnets in advanced undergraduate and applied graduate courses on variational inference, a popular optimization-based approach for approximate inference with probabilistic models. Our proposed module is guided by active learning principles: In addition to lecture materials on variational inference, we provide an accompanying class activity, an \texttt{R shiny} app, and guided labs based on real data applications of logistic regression and clustering documents using Latent Dirichlet Allocation with \texttt{R} code. The main goal of our module is to expose students to a method that facilitates statistical modeling and inference with large datasets. Using our proposed module as a foundation, instructors can adopt and adapt it to introduce more realistic case studies and applications in data science, Bayesian statistics, multivariate analysis, and statistical machine learning courses.
\end{abstract}

\noindent%
\vfill

\newpage
\spacingset{1.45} 

\newcommand\numberthis{\addtocounter{equation}{1}\tag{\theequation}}

\newcommand{\argmin}{\mathop{\mathrm{arg\,min}}}
\newcommand{\argmax}{\mathop{\mathrm{arg\,max}}}

\newcommand{\thetab}{\bm{\theta}}  
\newcommand{\Thetab}{\bm{\Theta}}
\newcommand{\yb}{\bm{y}}    
\newcommand{\Yb}{\bm{Y}}
\newcommand{\lambdab}{\bm{\lambda}}
\newcommand{\phib}{\bm{\phi}}
\newcommand{\Phib}{\bm{\Phi}}
\newcommand{\VK}{\textcolor{blue}}
\newcommand{\MH}{\textcolor{red}}

\hypertarget{intro}{%
\section{Introduction}\label{intro}}

With the recent and rapid expansion of both undergraduate and graduate curricula with offerings in data science, Bayesian statistics, multivariate data analysis, and statistical machine learning, probabilistic models and Bayesian methods have grown to become more popular \citep[\citet{Dogucu22}]{Mccoy20}. In many settings, a central task in applications of probabilistic models is the evaluation of posterior distribution \(p(\bm{\theta}\mid \bm{y})\) of \(m\) model parameters \(\bm{\theta}\in \mathbb{R}^{m}\) (\(m \geq 1\)) conditioned on the observed data \(\bm{y}= (y_1, \dots, y_n)\) provided by the Bayes' theorem
\begin{equation}
    p(\bm{\theta}\mid \bm{y}) = \frac{p(\bm{y}\mid \bm{\theta})p(\bm{\theta})}{p(\bm{y})} \propto p(\bm{y}\mid \bm{\theta})p(\bm{\theta}).
\end{equation}
Here, \(p(\bm{y}\mid \bm{\theta})\) is the sampling density given by the underlying
probabilistic model for data, \(p(\bm{\theta})\) is the prior density that represents our prior beliefs about \(\bm{\theta}\) before seeing the data, and \(p(\bm{y})\) is the marginal data distribution. The posterior distribution \(p(\bm{\theta}\mid \bm{y})\), however, has closed form only in a limited number of scenarios (e.g., conjugate priors) and therefore typically requires approximation. By far the most popular approximation methods are Markov chain Monte Carlo (MCMC) algorithms including Gibbs sampler, Metropolis, Metropolis-Hastings, and Hamiltonian Monte Carlo \citep{BDA}, to name a few. See \citet{alberthu2020} for a review of these algorithms in undergraduate Bayesian courses. While useful for certain scenarios, these MCMC algorithms do not scale well with large datasets and can have a hard time approximating multimodal posteriors \citep[\citet{bardenet2017}]{rudoy2006}. Such challenges therefore limit the applications of probabilistic models that can be discussed in the classroom and restrict students' exposure to more realistic case studies that include applying neural networks, pattern recognition, and natural language processing to massive datasets.

Variational inference is an alternative to the sampling-based approximation via MCMC that approximates a target density through optimization. Statisticians and computer scientists (starting with \citet{Peterson}, \citet{Jordan1999}, \citet{blei17}) have been using variational techniques in a variety of settings because these techniques tend to be faster and easier to scale to massive datasets. Despite its popularity among statistics and data science practitioners, variational inference is rarely discussed, especially in undergraduate courses, as it is believed to be a too advanced topic \citep{Dogucu22}. With this in mind, we have developed a one-week course module that serves as a gentle introduction to this topic. The goal is to help instructors to introduce variational techniques in their advanced undergraduate and applied graduate courses for more realistic case studies of probabilistic models. Our proposed one-week module is based on the best practices of active learning, which have been shown to improve student learning and engagement \citep[\citet{freedman14}, \citet{Kestin19}]{Michael06}. Our main guiding principle in designing the module is to involve students in the learning process by introducing student-centered class activities and labs. The guiding principle also includes assigning open-ended questions, focusing on problem-solving, providing appropriate scaffolding for activities, and creating opportunities to work collaboratively with peers.

Our module is designed for students to gain a fundamental understanding and practical experience with variational inference over the course of two class meetings. During the first meeting, students are exposed to the fundamentals of variational inferences including the Kullback-Leibler divergence, evidence lower bound, gradient ascent, and coordinate ascent. Additionally, they gain their first hands-on experience by applying variational inference to a simple probabilistic model for count data. To encourage and empower instructors to adopt and adapt this variational inference module, we provide an accompanying in-class handout and an \texttt{R Shiny} app with details in the supplementary materials. During the second class meeting, students work on a guided \texttt{R} lab to apply variational inference to a realistic scenario. We offer two lab options for instructors to choose from depending on the course level and student background. For advanced undergraduate courses, we provide a case study of U.S. women labor participation with logistic regression model. For more advanced and self-motivated undergraduate students and applied graduate students, we present an application of variational inference to clustering documents with Latent Dirichlet Allocation \citep{LDA2003}. See Table \ref{tab:module} for the breakdown of the module.

\begin{table}[h!]
\centering
\begin{tabular}{l|l|}
\cline{2-2}
 & \textbf{Content} \\ \hline
\multicolumn{1}{|l|}{\multirow{2}{*}{$1^{\text{st}}$ class}} & Lecture: Fundamentals of variational inference \\
\multicolumn{1}{|l|}{} & Class activity: Probabilistic model for count data with variational inference \\ \hline
\multicolumn{1}{|l|}{$2^{\text{nd}}$ class} & Lab: Logistic regression/Document clustering \\ \hline
\end{tabular}
\caption{\label{tab:module} Outline of the one-week variational inference module.}
\end{table}

As for the audience, we believe that the module can be seamlessly integrated into any advanced undergraduate or applied graduate course in data science, Bayesian statistics, multivariate data analysis, and statistical machine learning that covers topics on clustering, classification, or text analysis. The prerequisites needed for the module are a basic understanding of statistical modeling, probability distributions, and elementary calculus.

The remainder of the paper is organized as the following. In Section \ref{VI}, we provide an overview of variational inference essentials that can be readily used as a basis for a lecture instruction. Section \ref{example} presents a motivating example and the Gamma-Poisson model for count data that serves as the first hands-on class activity with variational inference. In Section \ref{application}, we offer realistic case studies for variational inference with implementation details in \texttt{R}, which can be used as a computing lab. We end the paper in Section \ref{conclusion} with a few concluding remarks.

\hypertarget{VI}{%
\section{Lecture: Foundations of Variantional Inference}\label{VI}}

In this section, we introduce concepts and definitions of variational inference in Section \ref{VI-basic}, discuss the choices of variational families in Section \ref{VI-family}, and present details of ELBO optimization in Section \ref{VI-ELBO}. We also include recommendations of variational families and ELBO optimization strategies with pedagogical considerations for an advanced undergraduate and applied graduate audience. Instructors can design their lecture based on these materials tailored to their needs.

\hypertarget{VI-basic}{%
\subsection{Concepts and Definitions}\label{VI-basic}}

The main idea behind variational inference is to approximate the target probability density \(p(\bm{\theta}\mid \bm{y})\) by a member of some relatively simple family of densities \(q(\bm{\theta}\mid \bm{\lambda})\), indexed by the variational parameter \(\bm{\lambda}\in \mathbb{R}^{m}\) (\(m \geq 1\)), over the space of model parameters \(\bm{\theta}\). Note that \(\bm{\lambda}=(\lambda_1, \dots, \lambda_m)\) has \(m\) components of (potentially) varying dimensions. Variational approximation is done by finding the member of variational family that minimizes the Kullback-Leibler (KL) divergence of \(q(\bm{\theta}\mid \bm{\lambda})\) from \(p(\bm{\theta}\mid \bm{y})\):
\begin{equation}\label{eqn:VI_def}
q^* = \mathop{\mathrm{arg\,min}}_{q(\bm{\theta}\mid \bm{\lambda})} KL(q(\bm{\theta}\mid \bm{\lambda})||p(\bm{\theta}\mid \bm{y})),
\end{equation}
with KL divergence being the expectation of the log ratio between the \(q(\bm{\theta}\mid \bm{\lambda})\) and \(p(\bm{\theta}\mid \bm{y})\) with respect to \(q(\bm{\theta}\mid \bm{\lambda})\):
\begin{equation}\label{eqn:KL_def}
 KL(q(\bm{\theta}\mid \bm{\lambda})||p(\bm{\theta}\mid \bm{y}))  = \mathbb{E}_{q}\big[\log \frac{q(\bm{\theta}\mid \bm{\lambda})}{p(\bm{\theta}\mid \bm{y})}\big] = \mathbb{E}_{q}\big[\log q(\bm{\theta}\mid \bm{\lambda})\big] - \mathbb{E}_{q}\big[\log p(\bm{y}, \bm{\theta})\big] + \log p(\bm{y}).
\end{equation}
The KL divergence measures how different is the probability distribution \(q(\bm{\theta}\mid \bm{\lambda})\) from \(p(\bm{\theta}\mid \bm{y})\) \citep{KL51}. Note that while we use the KL divergence to measure the similarity between two densities, it is not a metric because the KL divergence is not symmetric and does not satisfy the triangle inequality. In fact, the order of \(q(\bm{\theta}\mid \bm{\lambda})\) and \(p(\bm{\theta}\mid \bm{y})\) in Equation \eqref{eqn:VI_def} is deliberate as it leads to taking the expectation with respect to the
variational distribution \(q(\bm{\theta}\mid \bm{\lambda})\). One can naturally think of reversing the roles of \(q(\bm{\theta}\mid \bm{\lambda})\) and \(p(\bm{\theta}\mid \bm{y})\). However, this leads to a ``different kind'' of variational inference called \textit{expectation propagation} (\citet{Minka2001}), which loses computational efficiency of variational inference defined in Equation \eqref{eqn:VI_def}.

\begin{figure}

{\centering \includegraphics[width=0.6\linewidth]{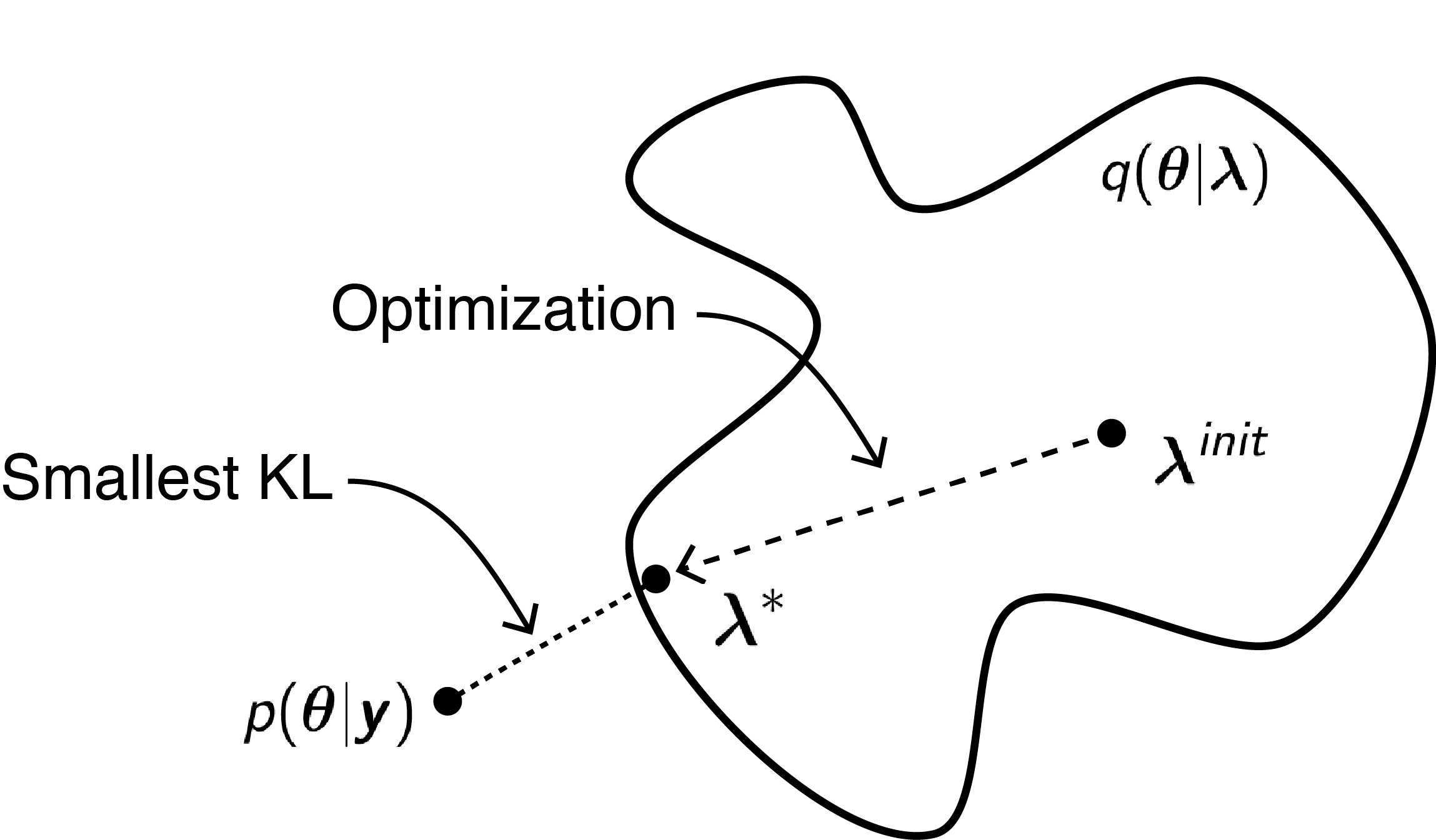} 

}

\caption{Illustration of variational inference as the optimization-based approximation. The goal of variational inference is to find a member of the variational family that minimizes KL divergence with the target distribution.}\label{fig:VI-idea}
\end{figure}

In a nutshell, rather than sampling, variational inference approximates densities using optimization. See Figure \ref{fig:VI-idea} for a graphical illustration, i.e., by finding the values of variational parameters from \(\bm{\lambda}^{init}\) to \(\bm{\lambda}^*\) through optimization which lead to a variational distribution \(q(\bm{\theta}\mid \bm{\lambda})\) that is close to the target posterior distribution \(p(\bm{\theta}\mid \bm{y})\) defined by the smallest KL divergence. Finding the optimal \(q^*\) is done in practice by maximizing an equivalent objective function, \(\mathcal{L}(\bm{\lambda})\), the \textit{evidence lower bound} (ELBO), because the KL divergence is intractable as it requires the evaluation of the marginal distribution \(p(\bm{y})\):
\begin{align*}
\mathcal{L}(\bm{\lambda}) &= \quad \quad \mathbb{E}_q[\log p(\bm{y}, \bm{\theta}) - \log q(\bm{\theta}| \bm{\lambda}) ]\\
& = \underbrace{\mathbb{E}_q[\log p(\bm{y}|\bm{\theta})]}_\text{Expected log-likelihood of data} - \underbrace{KL(q(\bm{\theta}|\bm{\lambda})||p(\bm{\theta}))}_\text{KL div. between the variational and prior densities}. \addtocounter{equation}{1}\tag{\theequation}\label{eqn:VBI_ELBO}
\end{align*}
Starting with Equation \eqref{eqn:KL_def}, one can derive the ELBO as the sum between the negative KL divergence of the variational density from the target density
and the log of the marginal density \(p(\bm{y})\). Since the term \(\log p(\bm{y})\) is constant with respect to \(q(\bm{\theta}\mid \bm{\lambda})\), the objective functions in Equation \eqref{eqn:KL_def} and Equation \eqref{eqn:VBI_ELBO} are equivalent. Examining the ELBO also reveals the intuition behind variational inference. On the one hand, the first term in Equation \eqref{eqn:VBI_ELBO} encourages the variational approximation to place mass on parameter values that maximize the sampling density \(p(\bm{y}\mid \bm{\theta})\). On the other hand, the second term in Equation \eqref{eqn:VBI_ELBO} prefers closeness of the variational density to the prior. Therefore, the ELBO shows a similar tension between the sampling density and the prior known in Bayesian inference.

\hypertarget{VI-family}{%
\subsection{Variational Families with Pedagogical Recommendations}\label{VI-family}}

We now move on to the implementation details of variational inference starting with the selection of the variational family \(q(\bm{\theta}\mid \bm{\lambda})\). This choice is crucial as it affects the complexity of optimization outlined in Section \ref{VI-basic} as well as the quality of variational approximation.

\hypertarget{mean-field-variational-family}{%
\subsubsection*{Mean-field Variational Family}\label{mean-field-variational-family}}
\addcontentsline{toc}{subsubsection}{Mean-field Variational Family}

\begin{figure}

{\centering \includegraphics[width=0.6\linewidth]{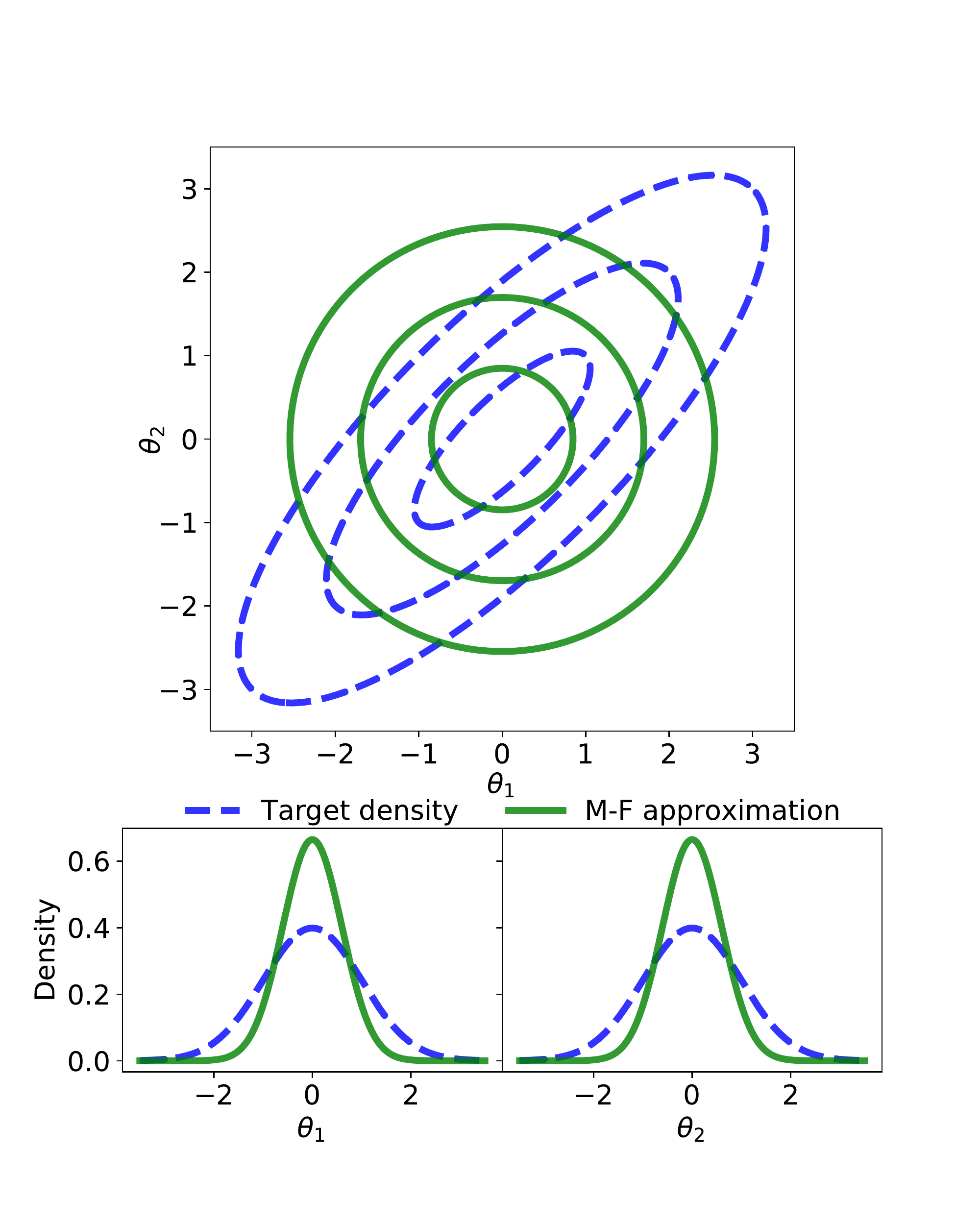} 

}

\caption{Mean-field variational approximation of a two-dimensional normal target density. The figure illustrates the common pitfall of the mean-field approximation in situations with correlated model parameters.}\label{fig:MF-approx}
\end{figure}

By far the most popular is the \textit{mean-field} variational family which assumes that all the unknown parameters are mutually independent, each approximated by its own univariate variational density:
\begin{equation}\label{eqn_q}
q(\bm{\theta}\mid \bm{\lambda}) = \prod_{i=1}^{m} q(\theta_i \mid \bm{\lambda}_i).
\end{equation}
For example, a typical choice for real-valued parameters is the normal variational family \(q(\theta \mid \mu, \sigma^2)\) and the log-normal or Gamma for non-negative parameters. The main advantage of the mean-field family is in its simplicity as it requires only a minimum number of parameters to be estimated (no correlation parameters) and often leads to uncomplicated optimization. However, the mutually independent parameter assumption comes at a price because the mean-field family cannot capture relationships between model parameters. To illustrate the pitfalls of mean-field approximation, consider a simple case of a two-dimensional normal target density with highly correlated components. Figure \ref{fig:MF-approx} shows the optimal mean-field variational approximation given by the product of two normal densities. One can clearly see that the optimal variational densities match well with the means of the target density, but the marginal variances are underestimated. To further understand this common flaw of mean-field approximation, consider the definition of KL divergence in Equation \eqref{eqn:KL_def}. The objective function penalizes more larger density in \(q(\bm{\theta}\mid \bm{\lambda})\) in areas where \(p(\bm{\theta}\mid \bm{y})\) has low density than the opposite direction (recall that the expectation is taken with respect to the variational density).

\hypertarget{recommendation-for-instruction}{%
\subsubsection*{Recommendation for Instruction}\label{recommendation-for-instruction}}
\addcontentsline{toc}{subsubsection}{Recommendation for Instruction}

It is worth noting that the development of new variational families which improves on the trade-off between complexity and expressiveness of variational approximations has been a fruitful and active area of research. To keep the scope of the one-week variational inference module manageable to both the students and the instructors, we recommend solely focusing on the mean-field approximation. For interested students who want to explore further, we encourage the instructors to refer them to the recent work of \citet{pmlr-v130-ambrogioni21a} that provides a detailed discussion on many state-of-the-art variational families and their associated implementation challenges.

\hypertarget{VI-ELBO}{%
\subsection{ELBO Optimization with Pedagogical Recommendations}\label{VI-ELBO}}

Besides the choice of variational family, another key implementation detail to address is the way in which we find the member of the variational family that maximizes the ELBO. Since this is a fairly general optimization problem, one can in principle use any optimization procedure. In the variational inference literature, the coordinate ascent and the gradient ascent procedures are the most prominent and widely used (\citet{blei17}).

\hypertarget{coordinate-ascent}{%
\subsubsection*{Coordinate Ascent}\label{coordinate-ascent}}
\addcontentsline{toc}{subsubsection}{Coordinate Ascent}

The coordinate ascent approach is based on the simple idea that one can maximize ELBO, which is a multivariate function, by cyclically maximizing it along one direction at a time. Starting with initial values (denoted by superscript \(0\)) of the \(m\) variational parameters \(\bm{\lambda}^0\)
\begin{equation*}
    \bm{\lambda}^0 = (\lambda^0_1, \dots,\lambda^0_m),
\end{equation*}
one obtains the \((k+1)^{\text{th}}\) updated value of variational parameters by iteratively solving
\begin{equation*}
    \lambda_i^{k + 1} = \mathop{\mathrm{arg\,max}}_{x} \mathcal{L}(\lambda_1^{k + 1}, \dots, \lambda_{i-1}^{k + 1}, x, \lambda_{i + 1}^k, \dots, \lambda_m^k),
\end{equation*}
which can be accomplished without using gradients \citep{blei17}.

\hypertarget{gradient-ascent}{%
\subsubsection*{Gradient Ascent}\label{gradient-ascent}}
\addcontentsline{toc}{subsubsection}{Gradient Ascent}

Variational inference via gradient ascent uses the standard iterative optimization algorithm based on the idea that the ELBO grows fastest in the direction of its gradient \citep{Hoffman13a}. In particular, the update of variational parameters \(\bm{\lambda}\) at the \((k+1)^{\text{th}}\) iteration is given by
\begin{equation*}
    \bm{\lambda}^{k+1} \leftarrow \bm{\lambda}^{k} + \eta \times \nabla_{\bm{\lambda}} \mathcal{L}(\bm{\lambda}^{k}),
\end{equation*}
where \(\nabla_{\bm{\lambda}} \mathcal{L}(\bm{\lambda})\) is the ELBO gradient, and \(\eta\) is the step size which is also called the learning rate. The step size controls the rate at which one updates the variational parameters.

For both coordinate and gradient ascent, we typically declare convergence of variational parameters once the change in ELBO falls below some small threshold \citep{blei17}.

\hypertarget{recommendation-for-instruction-1}{%
\subsubsection*{Recommendation for Instruction}\label{recommendation-for-instruction-1}}
\addcontentsline{toc}{subsubsection}{Recommendation for Instruction}

Our recommendation for this variational inference module is to take the route of gradient ascent. This pedagogical choice is guided by our combined experience of teaching statistical modeling, Bayesian statistics, and data science at various undergraduate levels to students with diverse statistical backgrounds. Our recommendation has also taken into account the pedagogical advantages and disadvantages of gradient ascent and coordinate ascent for the target audience: Variational inference via coordinate ascent, while conceptually straightforward, requires non-trivial and model-specific derivations which can easily obscure the overall goal of this one-week module to expand students' exposure to the state-of-the-art approximate inference for probabilistic models; gradient-based variational inference, in contrast, leads to a black-box optimization that does not require any model-specific derivations due to an extensive autodifferentiation capabilities of modern statistical software such as \texttt{RStan} \citep{stan_development_team_stan_2012} and \texttt{Python} packages \texttt{PyTorch} \citep{Torch} and \texttt{TensorFlow} \citep{tensorflow2015-whitepaper}, to name a few.

We believe that from an advanced undergraduate- and applied graduate-level pedagogical perspective, gradient descent reflects better the current data science pipeline and allows the instruction to be focused on conceptual understanding of variational inference rather than technical details. Of course, using gradient-based optimization requires the students to be familiar with partial derivatives. Such a pre-requisite potentially restricts the audience for our module to a course with a multivariable calculus prerequisite. Nevertheless, we believe that an instructor with sufficient preparation can explain the basics behind gradient ascent to an audience with a minimal calculus background.

\hypertarget{example}{%
\section{Class Activity: A Probabilistic Model for Count Data with Variational Inference}\label{example}}

In this section, we provide a fully developed hands-on class activity with variational inference for count data. Starting with a motivating example in Section \ref{example-motivation}, we give an overview of the Gamma-Poisson model in Section \ref{example-overview}, and discuss details of the variational inference of this model in Section \ref{example-VI}, illustrated with an \texttt{R Shiny} app we have developed for instruction purpose. Instructors can adopt and adapt this class activity based on these materials tailored to their needs.

\hypertarget{example-motivation}{%
\subsection{A Motivating Example}\label{example-motivation}}

To illustrate how ELBO optimization leads to a good approximation of target posterior distribution, we consider Poisson sampling with a Gamma prior, which is a popular one-parameter model for count data \citep[\citet{AlbertHu2019book}, \citet{bayesrules}]{BDA}. To get started, we provide the following motivating example:

\begin{quote}
    \textit{Our task is to estimate the average number of active users of a popular massively multiplier online role-playing game (mmorpg) playing between the peak evening hours 7 pm and 10 pm. This information can help game developers in allocating server resources and optimizing user experience. To estimate the average number of active users, we will consider the counts (in thousands) of active players collected during the peak evening hours over a two-week period in the past month.}
\end{quote}

We have chosen the Gamma-Poisson model as the probabilistic model in this class activity for two reasons. First, the Gamma-Poisson model is relatively easy to understand for students with an elementary knowledge of probability distributions. Second, the Gamma is a conjugate prior for Poisson sampling which means that one can derive the exact posterior distribution (another Gamma) and check the fidelity of variational approximation by comparing to the analytical Gamma solution. The learning objective of this class activity is to get students familiarized with various aspects of variational inference presented in Section \ref{VI}, such as ELBO and variational family, with a simple example. Afterwards, students are better prepared to move on to more realistic scenarios described in Section \ref{application}.

\hypertarget{example-overview}{%
\subsection{Overview of the Gamma-Poisson Model}\label{example-overview}}

We now provide an overview of the Gamma-Poisson model which can be readily turned into a class lecture. Suppose \(\bm{y}= (y_1, \dots, y_n)\) represent the observed counts in \(n\) time intervals where the counts are independent, and each \(y_i\) follows a Poisson distribution with the same rate parameter \(\theta > 0\). The joint probability mass function of \(\bm{y}= (y_1, \dots, y_n)\) is
\begin{equation}\label{eqn:posson}
    p(\bm{y}\mid \theta) = \prod^{n}_{i=1}p(y_i \mid \theta) \propto \theta^{\sum^n_{i=1}y_i} e ^{-n\theta}.
\end{equation}
The posterior distribution for the rate parameter \(\theta\) is our inference target as \(\theta\) represents the expected number of counts that occurs during the given time intervals. Note that the Poisson sampling relies on several assumptions about the sampling process: One assumes that the time interval is fixed, the counts occurring during different time intervals are independent, and the rate \(\theta\) at which the counts occur is constant over time.

The Gamma-Poisson conjugacy states that if \(\theta\) follows a Gamma prior distribution with shape and rate parameters \(\alpha\) and \(\beta\), it can be shown that the posterior distribution \(p(\theta \mid \bm{y})\) will also have a Gamma density. Namely, if
\begin{equation}\label{eqn:gamma_prior}
    \theta\sim \textrm{Gamma}(\alpha,\beta),
\end{equation}
then
\begin{equation}\label{eqn:gamma_posterior}
\theta \mid \bm{y}\sim \textrm{Gamma}(\alpha+ \sum_{i=1}^n y_i, \beta + n).
\end{equation}

In other words, given \(\alpha\), \(\beta\), and \(\bm{y}\), one can derive the analytical solution to the posterior of \(p(\theta \mid \bm{y})\) and can subsequently sample from \(\textrm{Gamma}(\alpha+ \sum_{i=1}^n y_i, \beta + n)\) to get posterior samples of \(\theta\). While no approximation is needed, it serves as a good example of illustrating how variational inference works in such a setting and allows evaluations of the performance of variational inference.

\hypertarget{example-VI}{%
\subsection{Variational Inference of the Gamma-Poisson Model}\label{example-VI}}

Recall from Section \ref{VI} that variational inference approximates the (unknown) posterior distribution of a parameter by a simple family of distributions. In this Gamma-Poisson case, we will approximate the posterior distribution \(p(\theta \mid \bm{y})\) by a log-normal distribution with mean \(\mu\) and standard deviation \(\sigma\):
\begin{equation}\label{eqn:log_norm}
    q(\theta \mid \mu, \sigma) = \frac{1}{\theta \sigma \sqrt{2\pi}} e^{-\frac{(\ln{\theta} - \mu)^2}{2\sigma^2}}.
\end{equation}
The log-normal distribution is a continuous probability distribution of a random variable whose logarithm is normally distributed. It is a popular variational family for non-negative parameters because it can be expressed as a (continuously) transformed normal distribution, and therefore it is amenable to automatic differentiation. Automatic differentiation is a computation method for derivatives in computer programs that relies on the application of chain rule in differential calculus. It provides accurate and fast numerical derivative evaluations that leads to machine learning algorithms (such as variational inference) that do not require users to manually work out and code derivatives \citep[\citet{baydin18}]{Kucukelbir2017}.

\begin{figure}

{\centering \includegraphics[width=0.9\linewidth]{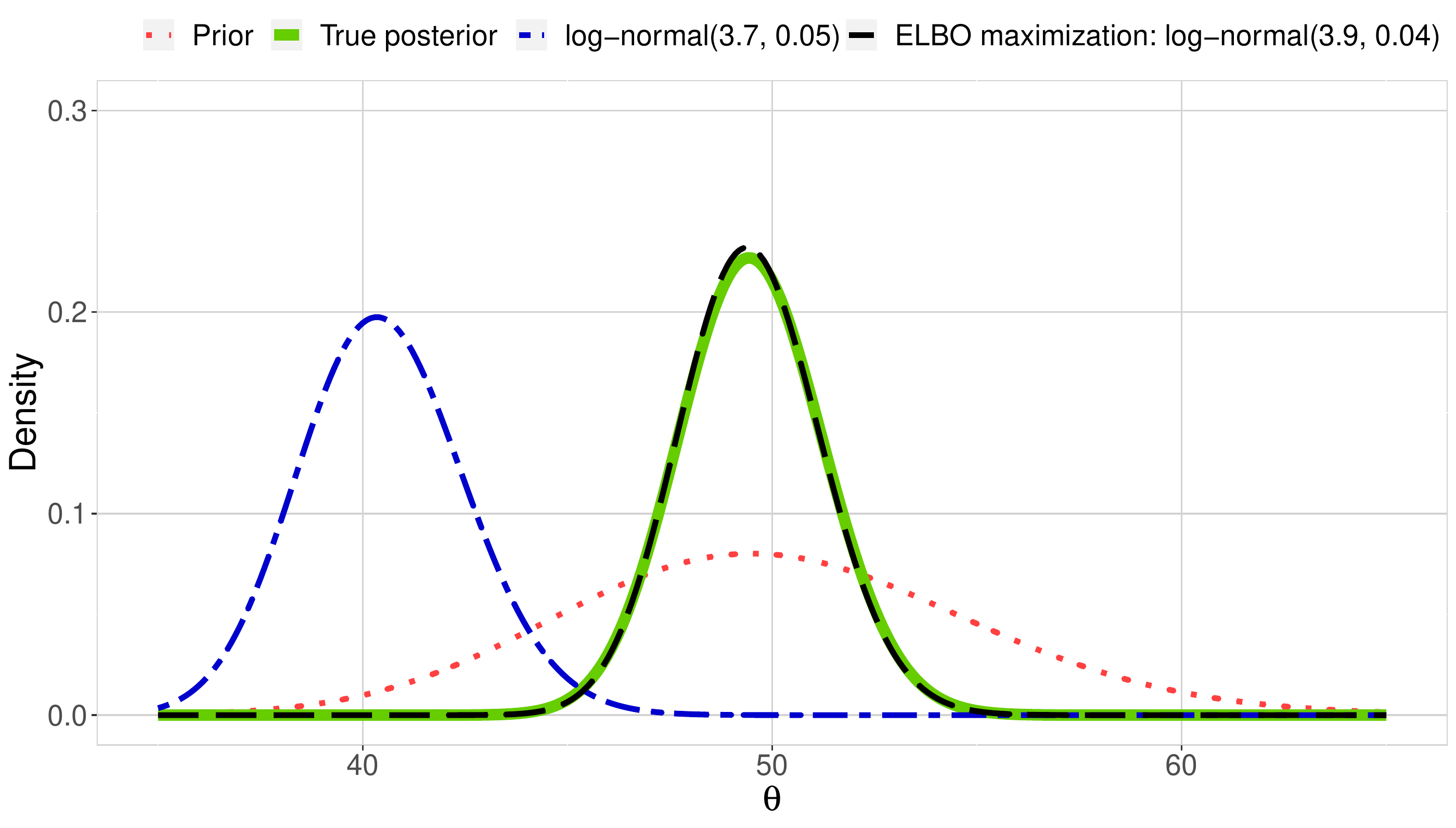} 

}

\caption{Variational approximation based on the motivating scenario of mmorpg's player activity. The true Gamma(792, 100) posterior and the prior Gamma(100,2) distributions are included. }\label{fig:VI-example}
\end{figure}

In the supplementary materials, we provide an accompanying in-class handout and an \texttt{R Shiny} app based on the motivating scenario of mmorpg described in Section \ref{example-motivation}. The first two parts of the handout present the motivating example and the overview of the Gamma-Poisson model. In the third part of the handout, students carry out exact posterior inference for the unknown rate parameter \(\theta\) using a small dataset of observed counts of mmorpg's active players. In the fourth and final part, students find variational approximation of \(p(\theta \mid \bm{y})\) and check how well their approximation matches the true posterior distribution. Figure \ref{fig:VI-example} shows the final variational approximation compared to the true Gamma(792, 100) posterior distribution from the handout example. We can see, on the one hand, the resulting log-normal(3.9, 0.04) distribution (the black dash line) that maximizes the ELBO visually overlaps with the true posterior (ELBO \(= -42.52\), KL divergence \(<0.001\)). On the other hand, another member of the variational family, the log-normal(3.7, 0.05) distribution (the blue dot-dash line; with ELBO \(= -57.55\) and KL divergence \(= 15.085\)), clearly differs from the target. This example illustrates the good performance of variational inference through optimization for the Gamma-Poisson count model.

The design of this class activity is guided by the active-learning principles listed in Section \ref{intro} and the goal is to give students their first hands-on experience with variational inference without the need of coding. Specifically, we include open-ended questions that focus on problem-solving and create opportunities for students to collaborate with peers. Moreover, the accompanying \texttt{R Shiny} app provides appropriate and sufficient scaffolding so that students can concentrate on conceptual understanding instead of the technical details, which follows our pedagogical recommendations in Section \ref{VI}.

We now turn to two guided \texttt{R} labs to illustrate the use of variational inference for more realistic case studies of logistic regression and document clustering.

\hypertarget{application}{%
\section{Labs: Logistic Regression and Document Clustering}\label{application}}

In what follows, we provide two alternatives for the lab portion of the proposed module. Section \ref{Logistic} outlines a case study of U.S. women labor participation with logistic regression model aimed for an advanced undergraduate audience. Section \ref{Logistic} applies variational inference on document clustering of a collection of Associate Press newspaper articles targeted for more advanced and self-motivated undergraduate students and students in applied graduate courses.

\hypertarget{Logistic}{%
\subsection{Logistic Regression}\label{Logistic}}

Logistic regression model is a popular supervised learning algorithm for binary classification due to its interpretability, solid predictive performance, and intuitive connection to the standard linear regression \citep{james2013introduction}. Despites its popularity, logistic regression, and its Bayesian version in particular, poses computational and statistical challenges in scenarios with large and high-dimensional data \citep{genkin2007}. For these reasons, we believe that a logistic regression is a suitable lab for an advanced (or an intermediate) undergraduate audience that can demonstrate the benefits of variational inference with a relatively low barrier from the statistical methodology point of view.

In Section \ref{Logistic-overview}, we briefly introduce the logistic regression model. In Section \ref{application-Labor}, we present a case study of U.S. women labor participation analysis where variational inference is implemented by the \texttt{cmdstanr R} package. We mainly focus on the interpretation of results and discuss pedagogical considerations and leave the details of the guided lab assignment with \texttt{R} code in the supplementary materials.

\hypertarget{Logistic-overview}{%
\subsubsection{Overview of Logistic Regression}\label{Logistic-overview}}

The logistic regression model assumes that a binary response \(y_i\) follows a Bernoulli distribution with probability of success \(p_i\):
\begin{equation*}
y_i \mid  p_i  \sim \textrm{Bernoulli}(p_i).
\end{equation*}

To relate a single predictor \(x_i\) to the response \(y_i\), logistic regression typically considers the natural logarithm of odds \(p_i / (1 - p_i)\) (also known as logit) to be a linear function of the predictor variable \(x_i\):
\begin{equation}
\textrm{logit}(p_i) = \textrm{ln} \bigg(\frac{p_i}{1-p_i}\bigg) = \alpha + \beta x_i,
\label{eq:lr}
\end{equation}
with \(\alpha\) and \(\beta\) being regression coefficients. Note that it is a bit more challenging to interpret the coefficients in the logistic regression than in standard linear regression as \(\alpha\) and \(\beta\) are directly related to the log odds \(p_i / (1 - p_i)\), instead of \(p_i\). For example, \(e^{\alpha}\) is the odds when the value of predictor \(x_i\) is 0, whereas the quantity \(e^{\beta}\) refers to the change in odds per unit increase in \(x_i\).

Lastly, by rearranging the terms in Equation \eqref{eq:lr}, one can express the probability of success \(p_i\) as
\begin{equation*}
p_i = \frac{e^{\alpha + \beta x_i}}{1 + e^{\alpha + \beta x_i}}.
\end{equation*}
In the Bayesian framework, one proceeds to prior specification of regression coefficients \((\alpha, \beta)\) and posterior inference through MCMC. For illustration, we consider independent normal priors for the regression coefficients \(\alpha \sim \textrm{Normal}(\mu_0, \sigma_0)\) and \(\beta \sim \textrm{Normal}(\mu_1, \sigma_1)\), where \((\mu_0, \mu_1)\) and \((\sigma_0, \sigma_1)\) are the prior means and standard deviations for the regression coefficients respectively.

\hypertarget{application-Labor}{%
\subsubsection{Predicting Labor Participation}\label{application-Labor}}

To apply variational inference in the context of logistic regression, we present a case study of predicting U.S. women labor participation. To do so, we consider a sample from the University of Michigan Panel Study of Income Dynamics (PSID), the longest running longitudinal household survey in the world. The survey dates back to 1968 and contains information on over 18,000 individuals living in 5,000 families in the United States. The survey of these individuals and their descendants has been collected continuously and includes data on income, wealth, employment status, health, marriage, and hundreds of other variables. Our interest is in analyzing a PSID sample of 753 observations from 1976 (\citet{Mroz1987}). The PSID 1976 survey is particularly interesting since it interviewed wives in households directly in the previous year. This PISD sample contains two variables: Family income exclusive of wife's income (in \$1000) and wife's labor participation (yes or no). The goal of the lab is predicting a wife's labor participation status (response variable \(y_i\)) from the family income exclusive of her income (predictor variable \(x_i\)) using logistic regression. We refer interested readers to \citet{AlbertHu2019book} Section 11.4 for an in-depth illustration of Bayesian logistic regression applied to the same prediction task.

\begin{figure}

{\centering \includegraphics[width=0.9\linewidth]{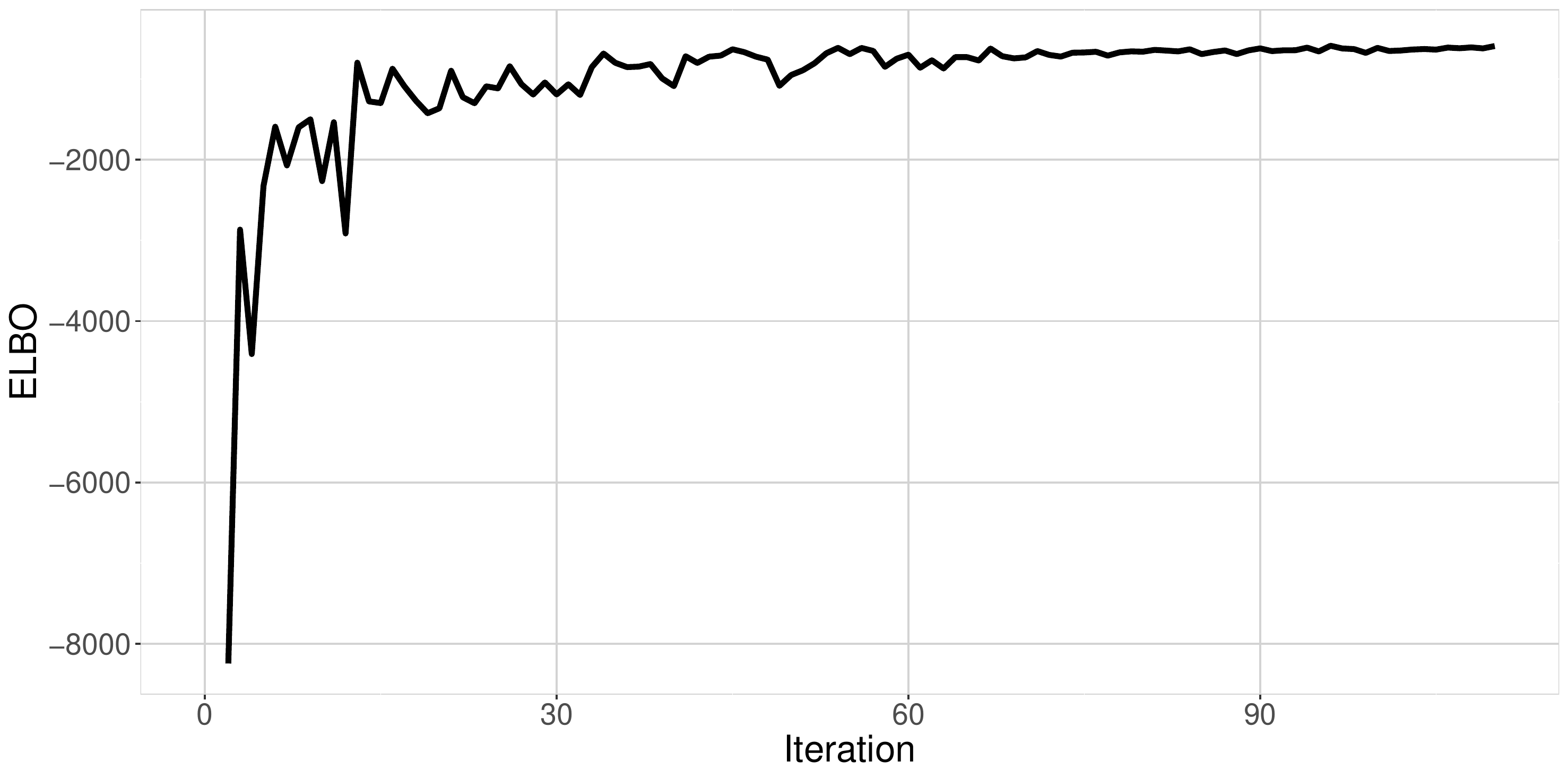} 

}

\caption{The evolution of ELBO for the logistic regression model based on a PSID sample of 753 observations from 1976.}\label{fig:Logistic-elbo}
\end{figure}

Figure \ref{fig:Logistic-elbo} shows the evolution of ELBO for the logistic regression model which converged after 80 iterations of the gradient ascent algorithm described in Section \ref{VI-ELBO}. We recommend running the algorithm repeatedly (i.e., 2-3 times) with a different random seed in the classroom and discussing the dependency of variational inference on initial values of variational parameters which can occur in practice. To highlight the computational benefits of variational inference, we also propose generating 50 replicates of the PSID sample (37,650 observations in total) and comparing the speed of convergence of variational approximation and MCMC approximation. The details of this exercise are provided in the supplementary materials.

\begin{figure}

{\centering \includegraphics[width=0.95\linewidth]{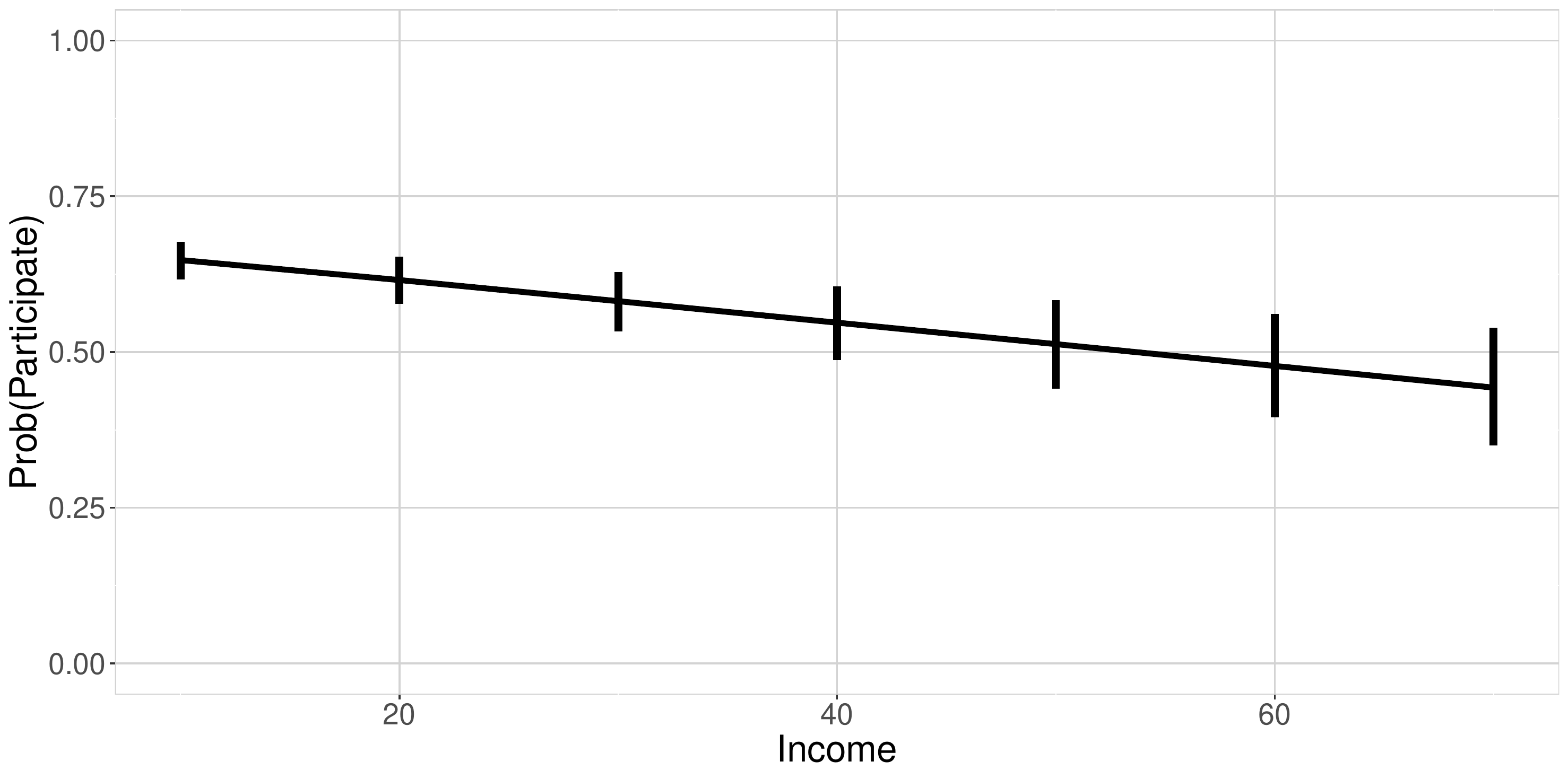} 

}

\caption{Posterior interval estimates for the probability of labor participation of a married woman who has a family income exclusive of her own income.}\label{fig:Logistic-pred}
\end{figure}

Figure \ref{fig:Logistic-pred} demonstrates one of the potential insights of the PSID survey data analysis, which is a series of posterior interval estimates for the probability of labor participation of a married woman who has a family income exclusive of her own income ranging from \$10,000 to \$70,000 with \$10,000 increments. One can see that in 1976, the likelihood of labor participation decreased with increasing family income exclusive of wife's income.

\hypertarget{LDA}{%
\subsection{Document Clustering}\label{LDA}}

Among the many models approximated by variational inference techniques, Latent Dirichlet Allocation (LDA) might be one of the most popular \citep{LDA2003}. LDA is a mixed-membership clustering model, commonly used for document clustering. Specifically, LDA models each document to have a mixture of topics, where each word in the document is drawn from a topic based on the mixing proportions \citep{stan_development_team_stan_2012}. While the LDA model is relatively easy and straightforward to follow, using conventional MCMC estimation techniques has proven to be too computationally demanding due to the large number of parameters involved. Therefore, researchers and practitioners turn to variational inference techniques when using LDA for document clustering \citep{LDA2003}.

In Section \ref{LDA-overview}, we briefly introduce the LDA model following the presentation in \citet{stan_development_team_stan_2012}. In Section \ref{application-APexample}, we present an LDA application to a collection of Associate Press newspaper articles where variational inference is implemented by the \texttt{cmdstanr R} package. For brevity, we focus on the interpretation of results and discuss pedagogical considerations and leave a \texttt{Stan} script for the LDA model and the details of the guided lab assignment with \texttt{R} code in the supplementary materials.

\hypertarget{LDA-overview}{%
\subsubsection{Overview of the LDA model}\label{LDA-overview}}

The LDA model considers \(K\) topics for \(M\) documents made up of words drawn from a vocabulary of \(V\) distinct words. For a document \(m\), a topic distribution \(\boldsymbol{\theta_m}\) over \(K\) topics is drawn from a Dirichlet distribution,
\begin{equation}
\boldsymbol{\theta}_m \sim \textrm{Dirichlet}(\boldsymbol{\alpha}),
\label{eq:theta}
\end{equation}
where \(\sum_{k=1}^{K}\theta_{m, k} = 1\) (\(0 \leq \theta_{m, k} \leq 1\)) and \(\boldsymbol{\alpha}\) is a vector of length \(K\) with positive values.

Each of the \(N_m\) words \(\{w_{m, 1},\dots, w_{m, N_m}\}\) in document \(m\) is then generated independently conditional on \(\boldsymbol{\theta}_m\). To do so, first, the topic \(z_{m, n}\) for word \(w_{m, n}\) in document \(m\) is drawn from
\begin{equation}
z_{m, n} \sim \textrm{categorical}(\boldsymbol{\theta}_m),
\label{eq:z}
\end{equation}
where \(\boldsymbol{\theta}_m\) is the document-specific topic-distribution defined in Equation \eqref{eq:theta}.

Next, the word \(w_{m, n}\) in document \(m\) is drawn from
\begin{equation}
w_{m, n} \sim \textrm{categorical}(\boldsymbol{\phi}_{z[m, n]}),
\label{eq:w}
\end{equation}
which is the word distribution for topic \(z_{m, n}\). Note that \(z[m, n]\) in Equation \eqref{eq:w} refers to \(z_{m, n}\).

Lastly, a Dirichlet prior is given to distributions \(\boldsymbol{\phi}_k\) over words for topic \(k\) as
\begin{equation}
\boldsymbol{\phi}_k \sim \textrm{Dirichlet}(\boldsymbol{\beta}),
\label{eq:phi}
\end{equation}
where \(\boldsymbol{\beta}\) is the prior a vector of length \(V\) (i.e., the total number of words) with positive values. Figure \ref{fig:LDA-model} shows a graphical model representation of LDA.

\begin{figure}

{\centering \includegraphics[width=0.9\linewidth]{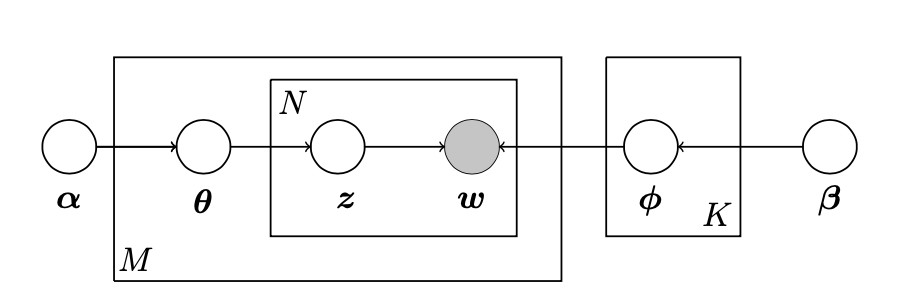} 

}

\caption{Graphical model representation of LDA. The largest box represents the documents. On the left, the inner box represents the topics and words within each document. On the right, the box represents the topics.}\label{fig:LDA-model}
\end{figure}

\hypertarget{application-APexample}{%
\subsubsection{Clustering of Associated Press Newspaper Articles}\label{application-APexample}}

As a realistic application of variational inference, we consider a collection of 2246 Associated Press newspaper articles to be clustered using the LDA model. The dataset is (conveniently) part of the \texttt{topicmodels R} package. We believe this dataset is well suited to demonstrate the capabilities of variational inference in the classroom as it is too large for the MCMC approximation to be feasible but small enough for the variational inference to take just a few minutes to converge. For brevity, we highlight the results based on a two-topic LDA model (i.e., \(K = 2\)) and leave the details to the guided lab in the supplementary materials. The number of topics is set to 2 for demonstration purposes and simplicity of interpretations. Comparing LDA with a different number of topics is often done with metrics such as semantic coherence or held-out data likelihood \citep{Mimno11}. While such a comparison is beyond the scope of this lab, interested students are encouraged to explore while being mentored by the instructors.

\begin{figure}

{\centering \includegraphics[width=0.9\linewidth]{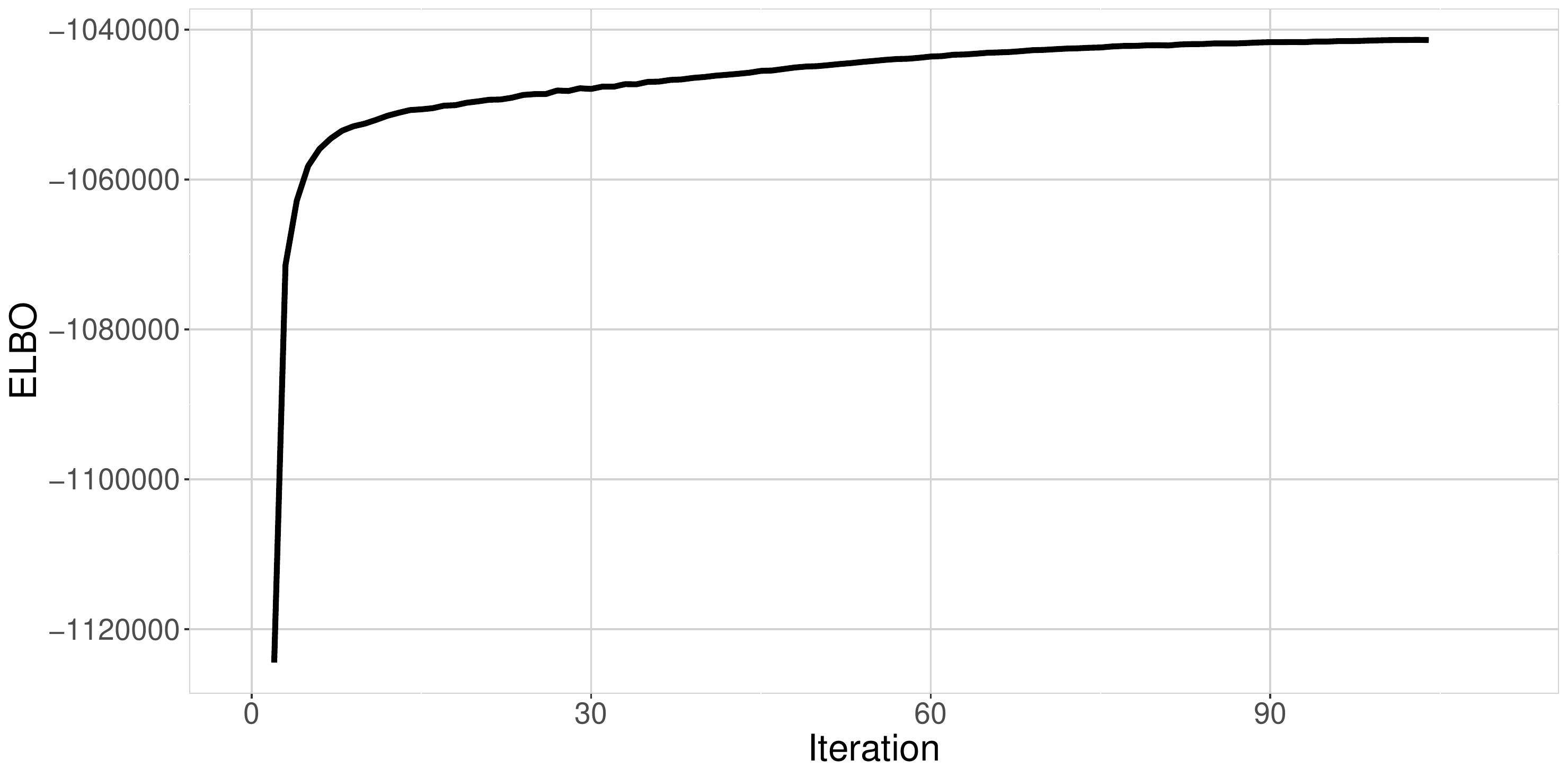} 

}

\caption{The evolution of ELBO for the two-topic LDA model based on 2246 Associated Press newspaper articles.}\label{fig:LDA-elbo}
\end{figure}

Figure \ref{fig:LDA-elbo} shows the evolution of ELBO for the two-topic LDA model which converged after a little bit over 100 iterations of the gradient ascent algorithm described in Section \ref{VI-ELBO}. On a standard laptop computer, this typically takes between 5-10 minutes depending on the CPU speed. Similar to the U.S. labor participation case study, we recommend running the algorithm repeatedly (i.e., 2-3 times) with a different random seed in the classroom and discussing the dependency of variational inference on initial values of variational parameters which can occur in practice.

\begin{figure}

{\centering \includegraphics[width=0.95\linewidth]{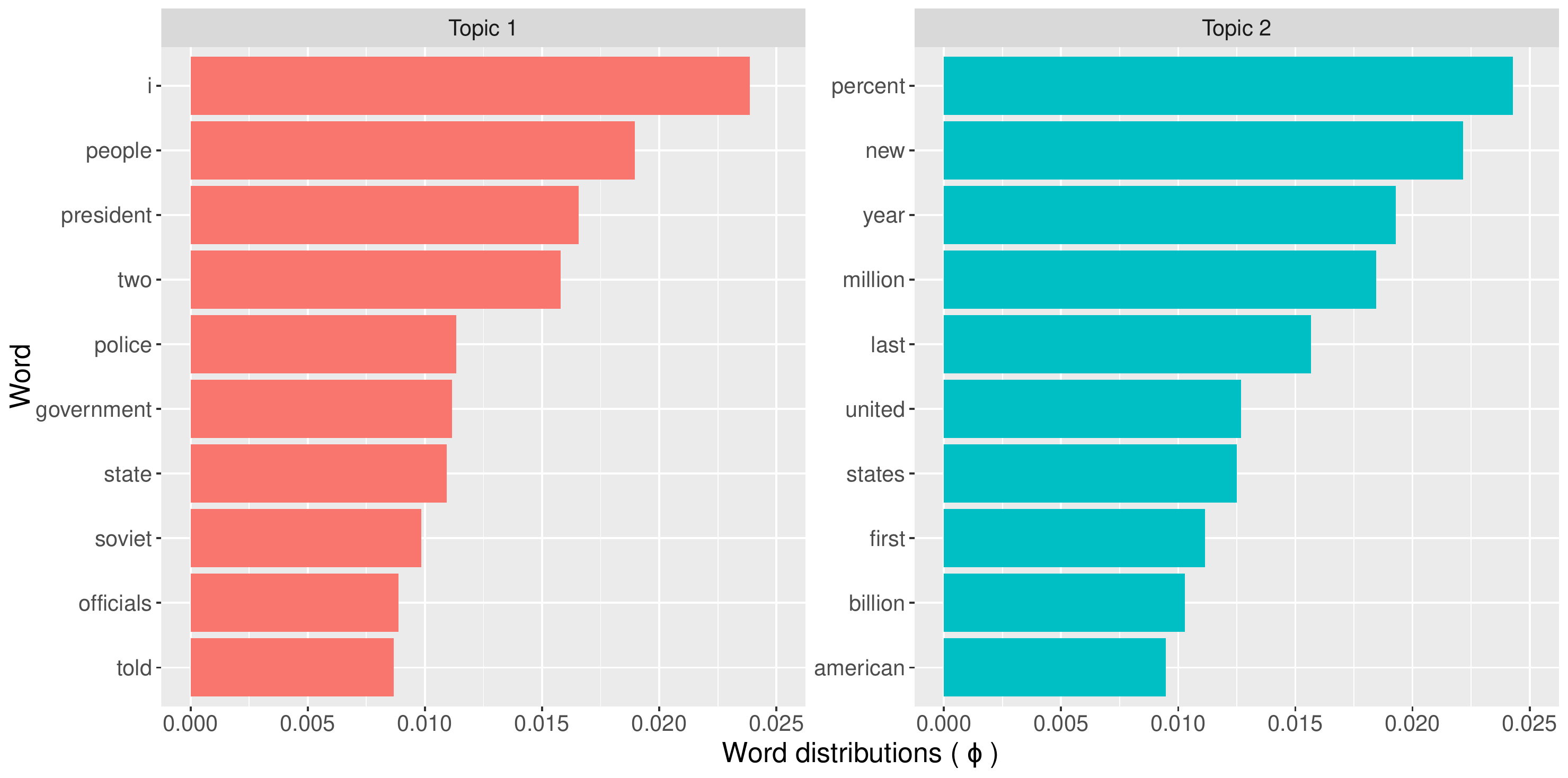} 

}

\caption{Word distributions based on the two-topic LDA model. The 10 most common words are displayed.}\label{fig:LDA-distr}
\end{figure}

\begin{figure}

{\centering \includegraphics[width=0.9\linewidth]{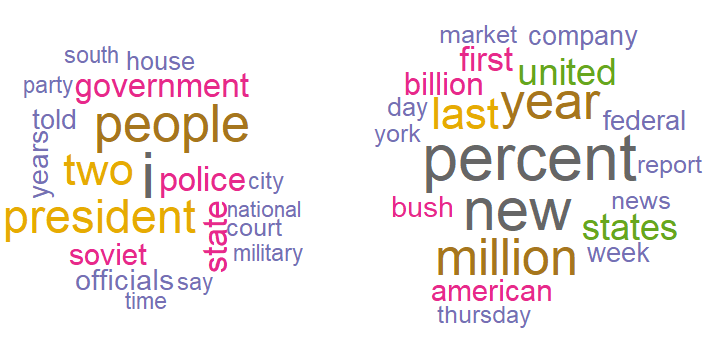} 

}

\caption{World clouds consisting of the 20 most common words for each of the two topics extracted by the LDA.}\label{fig:LDA-wordcloud}
\end{figure}

Figures \ref{fig:LDA-distr} and \ref{fig:LDA-wordcloud} are examples of graphical displays of the topics that were extracted from the collection of articles based with the LDA. In particular, Figure \ref{fig:LDA-distr} shows the 10 most common words for each topic; that is, the parts of distribution \(\boldsymbol{\phi}_k\), for \(k \in \{1,2\}\), with the largest mass. Figure \ref{fig:LDA-wordcloud} displays similar information for the 20 most common words for each topic in the form of a word cloud. The most common words in topic 1 include \textit{people, government, president, police,} and \textit{state}, suggesting that this topic may represent political news. In contrast, the most common words in topic 2 include \textit{percent, billion, million, market, American,} and \textit{states}, hinting that this topic may represent news about the US economy.

\hypertarget{conclusion}{%
\section{Concluding Remarks}\label{conclusion}}

In this paper, we present a newly-developed one-week course module that exposes students in advanced undergraduate and applied graduate courses to approximation via variational inference. The proposed module is self-contained in the sense that it encourages and empowers potential instructors to adopt and adapt the module as we provide an overview of variational inference, an active-learning-based class activity with an \texttt{R Shiny} app, and guided labs based on a realistic application with R code (see the supplementary materials or \url{https://github.com/kejzlarv/variational_inference_module}). Its design is rooted in the best practices of active learning that have been demonstrated to improve student learning and engagement.

The module can be integrated into any advanced undergraduate or applied graduate course where students learn probabilistic models (including logistic regression, Bayesian classifiers, neural networks, or models for natural language processing), such as Bayesian statistics, multivariate data analysis, and data science courses. The applications discussed in these courses are typically limited to scenarios with relatively small datasets, since the required use of MCMC does not scale well to large datasets. Given the popularity and scalability of variational inference, we hope that instructors adopting and adapting this module will be able to integrate more realistic and fun case studies in their classrooms. Moreover, the references and further readings provided in this paper are readily available resources for a deeper dive of variational inference by interested students with appropriate mentoring by their instructors.

\bibliographystyle{agsm}
\bibliography{bibliography.bib}

\newpage

{\bf Supplementary Materials for Introducing Variational Inference in Statistics and Data Science Curriculum}

The supplementary materials include: 1) Details of the class activity on probabilistic model for count data with variational inference, introduced in Section 3 in the main text; 2) The manual of the \texttt{R shiny} app we have developed for the module, mentioned in Section 3 in the main text; 3) Detais of the guided \texttt{R} logistic regression lab with U.S. women labor participation sample data, presented in Section 4.1 in the main text; and 4) Details of the guided \texttt{R} lab of the LDA application to a sample of the Associated Press newspaper articles with variational inference, presented in Section 4.2 in the main text.

\hypertarget{class-activity-probabilistic-model-for-count-data-with-variational-inference}{%
\section{Class Activity: Probabilistic Model for Count Data with Variational Inference}\label{class-activity-probabilistic-model-for-count-data-with-variational-inference}}

The goal of this activity is to illustrate variational inference on a simple example of Gamma-Poisson conjugate model, which is a popular model for count data.

\hypertarget{a-motivating-example}{%
\subsection{A Motivating Example}\label{a-motivating-example}}

Our task is to estimate the average number of active users of a popular massively multiplier online role-playing game (mmorpg) playing between the peak evening hours 7 pm and 10 pm. This information can help the game developers in allocating server resources and optimizing user experience. To make this estimate, we will consider the following counts (in thousands) of active players collected during the peak evening hours over a two-week period past month.

\begin{table}[h]
\begin{tabular}{l|lllllll|}
\cline{2-8}
 & \textbf{Sun} & \textbf{Mon} & \textbf{Tue} & \textbf{Wed} & \textbf{Thu} & \textbf{Fri} & \textbf{Sat} \\ \hline
\multicolumn{1}{|l|}{\textbf{Week 1}} & 50 & 47 & 46 & 52 & 49 & 55 & 53 \\ \hline
\multicolumn{1}{|l|}{\textbf{Week 2}} & 48 & 45 & 51 & 50 & 53 & 46 & 47 \\ \hline
\end{tabular}
\end{table}

\hypertarget{overview-of-the-gamma-poisson-model}{%
\subsection{Overview of the Gamma-Poisson Model}\label{overview-of-the-gamma-poisson-model}}

\textbf{Sampling density:}

Suppose that \(\bm{y}= (y_1, \dots, y_n)\) represent the observed counts in \(n\) time intervals where the counts are independent, then each \(y_i\) follows a Poisson distribution with rate \(\theta > 0\). Namely,
\[y_i \mid \theta \sim \textrm{Poisson}(\theta)\]

\begin{itemize}
\item $\mathbb{E}(y_i \mid \theta) = \theta$
\item $\mathbb{V}ar(y_i \mid \theta) = \theta$
\end{itemize}

\textbf{Prior distribution:}
\[\theta \sim \textrm{Gamma}(\alpha,\beta)\]

\begin{itemize}
\item $\alpha >0$ is the shape parameter
\item $\beta > 0$ is the rate parameter
\item $\mathbb{E}(\theta) = \frac{\alpha}{\beta}$
\item $\mathbb{V}ar(\theta) = \frac{\alpha}{\beta^2}$
\end{itemize}

\textbf{Posterior distribution:}
\[\theta \mid y_1, \dots, y_n \sim \textrm{Gamma}(\alpha+ \sum_{i=1}^n y_i, \beta + n)\]

\hypertarget{exact-inference-with-the-gamma-poisson-model}{%
\subsection{Exact Inference with the Gamma-Poisson Model}\label{exact-inference-with-the-gamma-poisson-model}}

We will start by selecting a prior distribution for the unknown average number of active users. Suppose that we elicit an expert's advice on the matter, and they tell us that a similar mmorpg has typically about 50,000 users during peak hours. However, they are not too sure about that, so the interval between 45,000 and 55,000 users should have a reasonably high probability. This reasoning leads to a \(\textrm{Gamma}(100,2)\) as a reasonable prior for the average number of active users.

\hypertarget{task-1-explain-the-reasoning-behind-using-textrmgamma1002-as-the-prior-distribution.}{%
\subsubsection*{\texorpdfstring{Task 1: Explain the reasoning behind using \(\textrm{Gamma}(100,2)\) as the prior distribution.}{Task 1: Explain the reasoning behind using \textbackslash textrm\{Gamma\}(100,2) as the prior distribution.}}\label{task-1-explain-the-reasoning-behind-using-textrmgamma1002-as-the-prior-distribution.}}
\addcontentsline{toc}{subsubsection}{Task 1: Explain the reasoning behind using \(\textrm{Gamma}(100,2)\) as the prior distribution.}

\hypertarget{task-2-use-the-information-above-to-find-the-exact-posterior-distribution-for-the-average-number-of-active-users.}{%
\subsubsection*{Task 2: Use the information above to find the exact posterior distribution for the average number of active users.}\label{task-2-use-the-information-above-to-find-the-exact-posterior-distribution-for-the-average-number-of-active-users.}}
\addcontentsline{toc}{subsubsection}{Task 2: Use the information above to find the exact posterior distribution for the average number of active users.}

\hypertarget{task-3-what-are-the-mean-and-standard-deviation-of-the-posterior-distribution-that-you-just-obtained-what-is-your-recommendation-about-the-typical-number-of-active-users-playing-the-mmorpg-between-the-peak-evening-hours-7pm-and-10pm}{%
\subsubsection*{Task 3: What are the mean and standard deviation of the posterior distribution that you just obtained? What is your recommendation about the typical number of active users playing the mmorpg between the peak evening hours 7pm and 10pm?}\label{task-3-what-are-the-mean-and-standard-deviation-of-the-posterior-distribution-that-you-just-obtained-what-is-your-recommendation-about-the-typical-number-of-active-users-playing-the-mmorpg-between-the-peak-evening-hours-7pm-and-10pm}}
\addcontentsline{toc}{subsubsection}{Task 3: What are the mean and standard deviation of the posterior distribution that you just obtained? What is your recommendation about the typical number of active users playing the mmorpg between the peak evening hours 7pm and 10pm?}

\newpage

\hypertarget{variational-inference-with-the-gamma-poisson-model}{%
\subsection{Variational Inference with the Gamma-Poisson Model}\label{variational-inference-with-the-gamma-poisson-model}}

Variational inference approximates the (unknown) posterior distribution of a parameter by a simple family of distributions. In this case, we will try to approximate the posterior distribution of the mmorpg's average number of active users between the peak hours \(\theta\) by a log-normal distribution with mean \(\mu\) and standard deviation \(\sigma\). Log-normal distribution is a continuous probability distribution of a random variable whose logarithm is normally distributed. It also happens to be a popular variational family for non-negative parameters as it is amenable to autodifferentiation. Since we know exactly how the posterior distribution for Gamma-Poisson model looks like, we will be able to check the fidelity of the variational approximation. Use the accompanying applet titled \textit{Variational Inference with Gamma-Poisson Model for count data} to complete the following task.

\hypertarget{task-4-use-the-sliders-in-the-applet-to-manually-find-the-member-of-a-log-normal-variational-family-that-well-approximates-the-posterior-distribution-of-theta.-what-is-your-strategy}{%
\subsubsection*{\texorpdfstring{Task 4: Use the sliders in the applet to manually find the member of a log-normal variational family that well approximates the posterior distribution of \(\theta\). What is your strategy?}{Task 4: Use the sliders in the applet to manually find the member of a log-normal variational family that well approximates the posterior distribution of \textbackslash theta. What is your strategy?}}\label{task-4-use-the-sliders-in-the-applet-to-manually-find-the-member-of-a-log-normal-variational-family-that-well-approximates-the-posterior-distribution-of-theta.-what-is-your-strategy}}
\addcontentsline{toc}{subsubsection}{Task 4: Use the sliders in the applet to manually find the member of a log-normal variational family that well approximates the posterior distribution of \(\theta\). What is your strategy?}

\hypertarget{task-5-compare-your-approximation-with-a-neighbor.-whose-approximation-is-closer-to-the-exact-posterior-distribution-of-theta-how-are-you-deciding}{%
\subsubsection*{\texorpdfstring{Task 5: Compare your approximation with a neighbor. Whose approximation is closer to the exact posterior distribution of \(\theta\)? How are you deciding?}{Task 5: Compare your approximation with a neighbor. Whose approximation is closer to the exact posterior distribution of \textbackslash theta? How are you deciding?}}\label{task-5-compare-your-approximation-with-a-neighbor.-whose-approximation-is-closer-to-the-exact-posterior-distribution-of-theta-how-are-you-deciding}}
\addcontentsline{toc}{subsubsection}{Task 5: Compare your approximation with a neighbor. Whose approximation is closer to the exact posterior distribution of \(\theta\)? How are you deciding?}

\hypertarget{task-6-check-the-box-in-the-applet-to-find-the-variational-approximation-using-the-gradient-ascent-algorithm.-how-close-was-the-variational-approximation-that-you-found-manually-to-the-one-found-here}{%
\subsubsection*{\texorpdfstring{Task 6: Check the \textit{Fit a variational approximation} box in the applet to find the variational approximation using the gradient ascent algorithm. How close was the variational approximation that you found manually to the one found here?}{Task 6: Check the  box in the applet to find the variational approximation using the gradient ascent algorithm. How close was the variational approximation that you found manually to the one found here?}}\label{task-6-check-the-box-in-the-applet-to-find-the-variational-approximation-using-the-gradient-ascent-algorithm.-how-close-was-the-variational-approximation-that-you-found-manually-to-the-one-found-here}}
\addcontentsline{toc}{subsubsection}{Task 6: Check the \textit{Fit a variational approximation} box in the applet to find the variational approximation using the gradient ascent algorithm. How close was the variational approximation that you found manually to the one found here?}

\newpage

\hypertarget{manual-of-the-app}{%
\section{\texorpdfstring{Manual of the \texttt{R shiny} app}{Manual of the  app}}\label{manual-of-the-app}}

\begin{figure}

{\centering \includegraphics[width=1\linewidth]{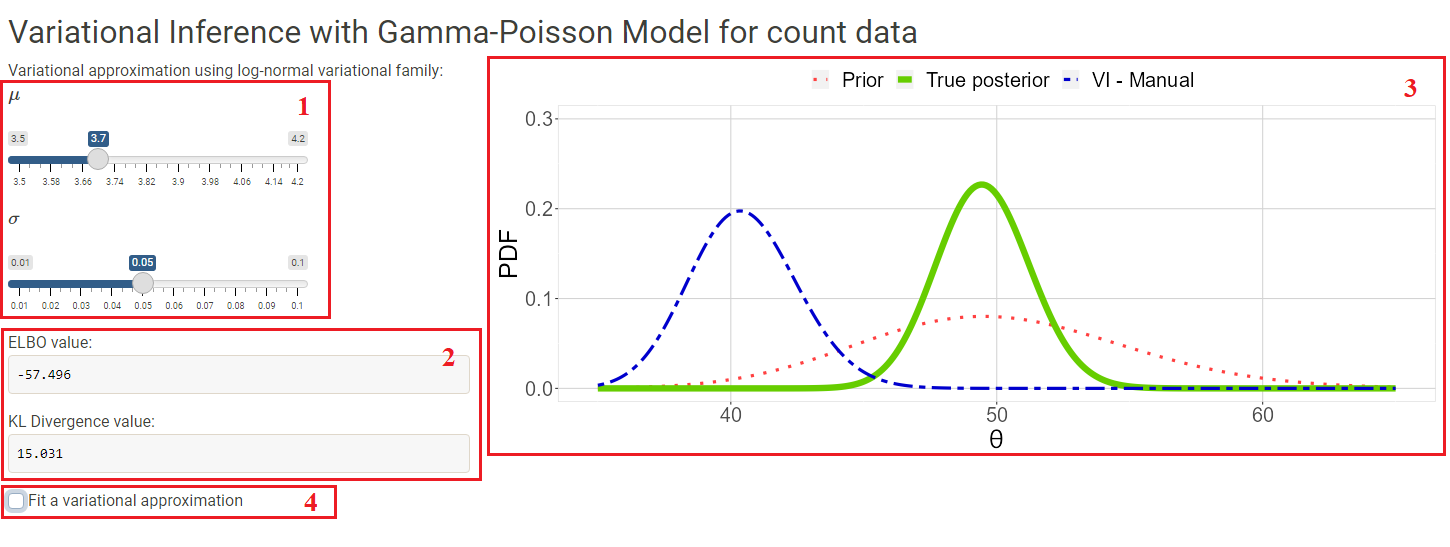} 

}

\caption{The applet is based on the class activity presented in Section 1 of the supplementary materials. The applet visual before checking the ``Fit a variational approximation`` checkbox is displayed.}\label{fig:app-manual-1}
\end{figure}

\begin{figure}

{\centering \includegraphics[width=1\linewidth]{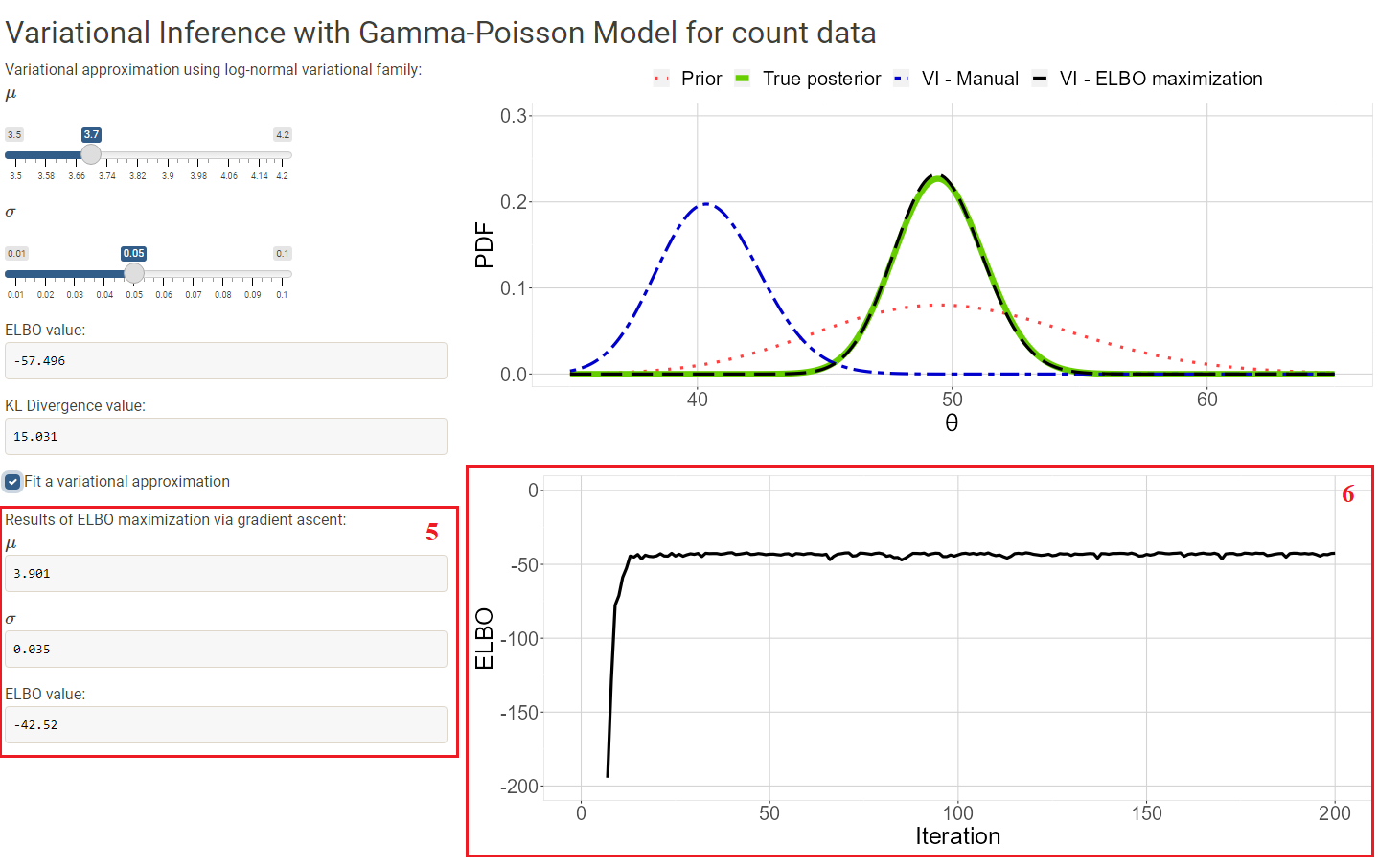} 

}

\caption{The applet visual after checking the ``Fit a variational approximation`` checkbox is displayed.}\label{fig:app-manual-2}
\end{figure}

This document describes the elements of \texttt{R Shiny} applet that accompanies the ``Probabilistic Model for Count Data with Variational Inference'' class activity. Note that the numbering in Section \ref{Manual1} and Section \ref{Manual2} corresponds to the numbered boxes in Figure \ref{fig:app-manual-1} and Figure \ref{fig:app-manual-2}.

\hypertarget{Manual1}{%
\subsection{Manual Search for Variational Approximation}\label{Manual1}}

\begin{enumerate}
\def\labelenumi{\arabic{enumi}.}
\item
  Sliders to control the mean \(\mu\) and the standard deviation \(\sigma\) of log-normal variational family.
\item
  The ELBO and KL divergence values for variational approximation based on the mean and standard deviations selected in box 1.
\item
  A plot that displays the true \(\textrm{Gamma}(792, 100)\) posterior distribution, the \(\textrm{Gamma}(100, 2)\) prior distribution, and the variational approximation based on the selection in box 1.
\item
  A checkbox to display the results of ELBO maximization via gradient ascent algorithm. The resulting variational approximation is plotted in box 3.
\end{enumerate}

\hypertarget{Manual2}{%
\subsection{Variational Approximation Based on ELBO Maximization}\label{Manual2}}

\begin{enumerate}
\def\labelenumi{\arabic{enumi}.}
\setcounter{enumi}{4}
\item
  The resulting mean \(\mu\), standard deviation \(\sigma\), and ELBO values of variational approximation based on ELBO maximization.
\item
  A plot depicting ELBO values for each iteration of the gradient ascent algorithm.
\end{enumerate}

\hypertarget{lab-on-logistic-regression}{%
\section{Lab on Logistic Regression}\label{lab-on-logistic-regression}}

The goal of this lab is to gain practical experience with variational inference on a case study of U.S. women labor participation with logistic regression model and implement the model in \texttt{R}.
To do so, we consider a sample data from the University of Michigan Panel Study of Income Dynamics (PSID) which is the longest running longitudinal household survey in the world. The survey dates back to 1968 and contains information on over 18,000 individuals living in 5,000 families in the United States. The survey of these individuals and their descendants has been collected continuously and includes data on income, wealth, employment status, health, marriage, and hundreds of other variables. Our interest is in analyzing a PSID sample of 753 observations from 1976 (\citet{Mroz1987}). The PSID 1976 survey is particularly interesting since it interviewed wives in households directly in the previous year. You can load the dataset \texttt{PSID} with the following R command.

\begin{Shaded}
\begin{Highlighting}[]
\NormalTok{PSID }\OtherTok{\textless{}{-}} \FunctionTok{read.csv}\NormalTok{(}\StringTok{"https://bit.ly/LaborVar"}\NormalTok{)}
\end{Highlighting}
\end{Shaded}

This PISD sample contains two variables: Family income exclusive of wife's income (in \$1000) and wife's labor participation (yes or no). The goal of the lab is predicting a wife's labor participation status (response variable) from the family income exclusive of her income (predictor variable) using logistic regression. We refer interested readers to \citet{AlbertHu2019book} Section 11.4 for an in-depth analysis of Bayesian logistic regression applied to the same prediction task.

\hypertarget{application-Logisticmodel}{%
\subsection{Overview of the Logistic Regression Model and Stan Script}\label{application-Logisticmodel}}

Logistic regression model is a supervised learning algorithm for binary classification. It is therefore well suitable for analysis of a binary response such as labor participation. Namely, the logistic regression model assumes that a binary response \(y_i\) follows a Bernoulli distribution with probability of success \(p_i\):
\begin{equation*}
y_i \mid  p_i  \sim \textrm{Bernoulli}(p_i).
\end{equation*}

To relate a predictor \(x_i\) to the response \(y_i\), logistic regression typically considers the natural logarithm of odds \(p_i / (1 - p_i)\) (also known as logit) to be a linear function of the predictor variables \(x_i\):
\begin{equation}
\textrm{logit}(p_i) = \textrm{ln} \bigg(\frac{p_i}{1-p_i}\bigg) = \alpha + \beta x_i,
\label{eq:lr}
\end{equation}
with \(\alpha\) and \(\beta\) being regression coefficients. Note that it is a bit more challenging to interpret the coefficients in the logistic regression than in standard linear regression as \(\alpha\) and \(\beta\) are directly related to the log odds \(p_i / (1 - p_i)\), instead of \(p_i\). For example, \(e^{\alpha}\) is the odds when the value of predictor \(x_i\) is , whereas the quantity \(e^{\beta}\) refers to the change in odds per unit increase in \(x_i\).

Lastly, by rearranging the terms in Equation \eqref{eq:lr}, one can express the probability of success \(p_i\) as
\begin{equation*}
p_i = \frac{e^{\alpha + \beta x_i}}{1 + e^{\alpha + \beta x_i}}.
\end{equation*}
In the Bayesian framework, one proceeds to prior specification of regression coefficients \((\alpha, \beta)\) and posterior inference through MCMC. For illustration, we consider independent normal priors for the regression coefficients \(\alpha \sim \textrm{Normal}(\mu_0, \sigma_0)\) and \(\beta \sim \textrm{Normal}(\mu_1, \sigma_1)\), where \((\mu_0, \mu_1)\) and \((\sigma_0, \sigma_1)\) are the prior means and standard deviations for the regression coefficients respectively.

Below, we include the \texttt{Stan} script for the logistic regression model of PSID data.

\begin{Shaded}
\begin{Highlighting}[]
\KeywordTok{data}\NormalTok{ \{}
  \DataTypeTok{int}\NormalTok{\textless{}}\KeywordTok{lower}\NormalTok{=}\DecValTok{0}\NormalTok{\textgreater{} N;}
  \DataTypeTok{vector}\NormalTok{[N] x;}
  \DataTypeTok{int}\NormalTok{\textless{}}\KeywordTok{lower}\NormalTok{=}\DecValTok{0}\NormalTok{,}\KeywordTok{upper}\NormalTok{=}\DecValTok{1}\NormalTok{\textgreater{} y[N];}
\NormalTok{\}}
\KeywordTok{parameters}\NormalTok{ \{}
  \DataTypeTok{real}\NormalTok{ alpha; }\CommentTok{// intercept}
  \DataTypeTok{real}\NormalTok{ beta; }\CommentTok{// slope}
\NormalTok{\}}
\KeywordTok{model}\NormalTok{ \{}
\NormalTok{  y \textasciitilde{} bernoulli\_logit(alpha + beta * x);}
  \CommentTok{// priors}
\NormalTok{  alpha \textasciitilde{} normal(}\DecValTok{0}\NormalTok{, }\DecValTok{5}\NormalTok{);}
\NormalTok{  beta \textasciitilde{} normal(}\DecValTok{0}\NormalTok{, }\DecValTok{5}\NormalTok{);}
\NormalTok{\}}
\end{Highlighting}
\end{Shaded}

\hypertarget{Lab-Logisticmodel}{%
\subsection{Variational Inference with the Logictic Regression Model}\label{Lab-Logisticmodel}}

We are now ready to fit the logistic regression model using variational inference capabilities of the \texttt{cmdstanr} package. The following code achieves the goal:

\begin{Shaded}
\begin{Highlighting}[]
\NormalTok{Logistic\_model\_cmd }\OtherTok{\textless{}{-}} \FunctionTok{cmdstan\_model}\NormalTok{(}\AttributeTok{stan\_file =} \StringTok{"Logistic.stan"}\NormalTok{)}

\NormalTok{data }\OtherTok{=} \FunctionTok{list}\NormalTok{(}\AttributeTok{N =} \FunctionTok{dim}\NormalTok{(PSID)[}\DecValTok{1}\NormalTok{],}
            \AttributeTok{x =}\NormalTok{ PSID}\SpecialCharTok{$}\NormalTok{FamilyIncome,}
            \AttributeTok{y =}\NormalTok{ PSID}\SpecialCharTok{$}\NormalTok{Participation}
\NormalTok{)}

\NormalTok{vi\_fit }\OtherTok{\textless{}{-}}\NormalTok{ Logistic\_model\_cmd}\SpecialCharTok{$}\FunctionTok{variational}\NormalTok{(}\AttributeTok{data =}\NormalTok{ data,}
                                    \AttributeTok{seed =} \DecValTok{1}\NormalTok{,}
                                    \AttributeTok{output\_samples =} \DecValTok{5000}\NormalTok{,}
                                    \AttributeTok{eval\_elbo =} \DecValTok{1}\NormalTok{,}
                                    \AttributeTok{grad\_samples =} \DecValTok{15}\NormalTok{,}
                                    \AttributeTok{elbo\_samples =} \DecValTok{15}\NormalTok{,}
                                    \AttributeTok{algorithm =} \StringTok{"meanfield"}\NormalTok{,}
                                    \AttributeTok{output\_dir =} \ConstantTok{NULL}\NormalTok{,}
                                    \AttributeTok{iter =} \DecValTok{1000}\NormalTok{,}
                                    \AttributeTok{adapt\_iter =} \DecValTok{20}\NormalTok{,}
                                    \AttributeTok{save\_latent\_dynamics=}\ConstantTok{TRUE}\NormalTok{,}
                                    \AttributeTok{tol\_rel\_obj =} \DecValTok{10}\SpecialCharTok{\^{}{-}}\DecValTok{4}\NormalTok{) }
\end{Highlighting}
\end{Shaded}

The ``Logistic.stan'' file contains the \texttt{Stan} script for the logistic regression model provided in Section \ref{application-Logisticmodel}. We recommend the usage of the \texttt{R} help to get familiar with the \texttt{variational()} method of the \texttt{cmdstan\_model()} function. The variable \texttt{vi\_fit} contains the results of variational approximation of the logistic regression parameters. For example, one can obtain the distribution of \(\beta\) with \texttt{vi\_fit\$summary("beta")}.

Finally, to access the ELBO values, use the following:

\begin{Shaded}
\begin{Highlighting}[]
\NormalTok{vi\_diag }\OtherTok{\textless{}{-}}\NormalTok{ utils}\SpecialCharTok{::}\FunctionTok{read.csv}\NormalTok{(vi\_fit}\SpecialCharTok{$}\FunctionTok{latent\_dynamics\_files}\NormalTok{()[}\DecValTok{1}\NormalTok{],}
                           \AttributeTok{comment.char =} \StringTok{"\#"}\NormalTok{)}
\NormalTok{ELBO }\OtherTok{\textless{}{-}} \FunctionTok{data.frame}\NormalTok{(}\AttributeTok{Iteration =}\NormalTok{ vi\_diag[,}\DecValTok{1}\NormalTok{], }\AttributeTok{ELBO =}\NormalTok{ vi\_diag[,}\DecValTok{3}\NormalTok{])}
\end{Highlighting}
\end{Shaded}

\hypertarget{task-1-use-a-graphical-display-to-show-the-posterior-interval-estimates-for-the-probability-of-labor-participation-of-a-married-woman-who-has-a-family-income-exclusive-of-her-own-income-ranging-from-10000-to-70000-with-10000-increments.}{%
\subsubsection*{Task 1: Use a graphical display to show the posterior interval estimates for the probability of labor participation of a married woman who has a family income exclusive of her own income ranging from \$10,000 to \$70,000 with \$10,000 increments.}\label{task-1-use-a-graphical-display-to-show-the-posterior-interval-estimates-for-the-probability-of-labor-participation-of-a-married-woman-who-has-a-family-income-exclusive-of-her-own-income-ranging-from-10000-to-70000-with-10000-increments.}}
\addcontentsline{toc}{subsubsection}{Task 1: Use a graphical display to show the posterior interval estimates for the probability of labor participation of a married woman who has a family income exclusive of her own income ranging from \$10,000 to \$70,000 with \$10,000 increments.}

\hypertarget{task-2-obtain-an-mcmc-fit-the-logistic-regression-model-and-compare-the-mcmc-approximation-with-the-variational-approximation.-what-do-you-observe}{%
\subsubsection*{Task 2: Obtain an MCMC fit the logistic regression model and compare the MCMC approximation with the variational approximation. What do you observe?}\label{task-2-obtain-an-mcmc-fit-the-logistic-regression-model-and-compare-the-mcmc-approximation-with-the-variational-approximation.-what-do-you-observe}}
\addcontentsline{toc}{subsubsection}{Task 2: Obtain an MCMC fit the logistic regression model and compare the MCMC approximation with the variational approximation. What do you observe?}

\hypertarget{task-3-create-a-dataset-that-contains-50-replicates-of-the-original-psid-dataset-37650-observations-in-total.-obtain-both-mcmc-and-variational-fit-for-the-regression-model-and-compare-the-times-that-it-takes-to-obtain-each-of-the-approximations.}{%
\subsubsection*{Task 3: Create a dataset that contains 50 replicates of the original PSID dataset (37,650 observations in total). Obtain both MCMC and variational fit for the regression model and compare the times that it takes to obtain each of the approximations.}\label{task-3-create-a-dataset-that-contains-50-replicates-of-the-original-psid-dataset-37650-observations-in-total.-obtain-both-mcmc-and-variational-fit-for-the-regression-model-and-compare-the-times-that-it-takes-to-obtain-each-of-the-approximations.}}
\addcontentsline{toc}{subsubsection}{Task 3: Create a dataset that contains 50 replicates of the original PSID dataset (37,650 observations in total). Obtain both MCMC and variational fit for the regression model and compare the times that it takes to obtain each of the approximations.}

All necessary R code for fitting the logistic regression model to the PSID sample, including the graphical displays shown in the main text, is included in a separate R script file called Logistic\_LAB.R available as a part of the supplementary materials. We also include a printout of the R script below for interested readers.

\begin{Shaded}
\begin{Highlighting}[]
\FunctionTok{library}\NormalTok{(cmdstanr)}
\CommentTok{\# Checking integrity of installation of cmdstanr}
\FunctionTok{check\_cmdstan\_toolchain}\NormalTok{()}
\FunctionTok{install\_cmdstan}\NormalTok{(}\AttributeTok{cores =} \DecValTok{2}\NormalTok{)}
\FunctionTok{cmdstan\_path}\NormalTok{()}
\FunctionTok{cmdstan\_version}\NormalTok{()}

\CommentTok{\# Auxiliary packages}
\FunctionTok{library}\NormalTok{(tidyverse)}

\DocumentationTok{\#\# Get data}
\NormalTok{PSID }\OtherTok{\textless{}{-}} \FunctionTok{read.csv}\NormalTok{(}\StringTok{"https://bit.ly/LaborVar"}\NormalTok{)}

\DocumentationTok{\#\# Input for stan model}
\NormalTok{data }\OtherTok{=} \FunctionTok{list}\NormalTok{(}\AttributeTok{N =} \FunctionTok{dim}\NormalTok{(PSID)[}\DecValTok{1}\NormalTok{],}
            \AttributeTok{x =}\NormalTok{ PSID}\SpecialCharTok{$}\NormalTok{FamilyIncome,}
            \AttributeTok{y =}\NormalTok{ PSID}\SpecialCharTok{$}\NormalTok{Participation}
\NormalTok{)}

\DocumentationTok{\#\# VI fit}
\NormalTok{Logistic\_model\_cmd }\OtherTok{\textless{}{-}} \FunctionTok{cmdstan\_model}\NormalTok{(}\AttributeTok{stan\_file =} \StringTok{"Logistic.stan"}\NormalTok{)}
\NormalTok{Logistic\_model\_cmd}\SpecialCharTok{$}\FunctionTok{print}\NormalTok{()  }

\NormalTok{vi\_fit }\OtherTok{\textless{}{-}}\NormalTok{ Logistic\_model\_cmd}\SpecialCharTok{$}\FunctionTok{variational}\NormalTok{(}\AttributeTok{data =}\NormalTok{ data,}
                                    \AttributeTok{seed =} \DecValTok{1}\NormalTok{,}
                                    \AttributeTok{output\_samples =} \DecValTok{5000}\NormalTok{,}
                                    \AttributeTok{eval\_elbo =} \DecValTok{1}\NormalTok{,}
                                    \AttributeTok{grad\_samples =} \DecValTok{15}\NormalTok{,}
                                    \AttributeTok{elbo\_samples =} \DecValTok{15}\NormalTok{,}
                                    \AttributeTok{algorithm =} \StringTok{"meanfield"}\NormalTok{,}
                                    \AttributeTok{output\_dir =} \ConstantTok{NULL}\NormalTok{,}
                                    \AttributeTok{iter =} \DecValTok{1000}\NormalTok{,}
                                    \AttributeTok{adapt\_iter =} \DecValTok{20}\NormalTok{,}
                                    \AttributeTok{save\_latent\_dynamics=}\ConstantTok{TRUE}\NormalTok{,}
                                    \AttributeTok{tol\_rel\_obj =} \DecValTok{10}\SpecialCharTok{\^{}{-}}\DecValTok{4}\NormalTok{) }

\DocumentationTok{\#\# Plotting ELBO}
\NormalTok{vb\_diag }\OtherTok{\textless{}{-}}\NormalTok{ utils}\SpecialCharTok{::}\FunctionTok{read.csv}\NormalTok{(vi\_fit}\SpecialCharTok{$}\FunctionTok{latent\_dynamics\_files}\NormalTok{()[}\DecValTok{1}\NormalTok{],}
                           \AttributeTok{comment.char =} \StringTok{"\#"}\NormalTok{)}
\NormalTok{ELBO }\OtherTok{=} \FunctionTok{data.frame}\NormalTok{(}\AttributeTok{Iteration =}\NormalTok{ vb\_diag[,}\DecValTok{1}\NormalTok{], }\AttributeTok{ELBO =}\NormalTok{ vb\_diag[,}\DecValTok{3}\NormalTok{])}

\FunctionTok{ggplot}\NormalTok{(}\AttributeTok{data =}\NormalTok{ ELBO, }\FunctionTok{aes}\NormalTok{(}\AttributeTok{x =}\NormalTok{ Iteration, }\AttributeTok{y =}\NormalTok{ ELBO)) }\SpecialCharTok{+} \FunctionTok{geom\_line}\NormalTok{(}\AttributeTok{lwd=}\FloatTok{1.5}\NormalTok{) }\SpecialCharTok{+} 
  \FunctionTok{theme}\NormalTok{(}\AttributeTok{text =} \FunctionTok{element\_text}\NormalTok{(}\AttributeTok{size =} \DecValTok{20}\NormalTok{),}
        \AttributeTok{panel.background =} \FunctionTok{element\_rect}\NormalTok{(}\AttributeTok{fill =} \StringTok{"transparent"}\NormalTok{,}
                                        \AttributeTok{color =} \StringTok{"lightgrey"}\NormalTok{),}
        \AttributeTok{panel.grid.major =} \FunctionTok{element\_line}\NormalTok{(}\AttributeTok{colour =} \StringTok{"lightgrey"}\NormalTok{)) }\SpecialCharTok{+}
  \FunctionTok{xlim}\NormalTok{(}\DecValTok{0}\NormalTok{,}\DecValTok{110}\NormalTok{)}

\DocumentationTok{\#\# Accessing parameters}
\NormalTok{vi\_fit}\SpecialCharTok{$}\FunctionTok{summary}\NormalTok{(}\StringTok{"alpha"}\NormalTok{) }
\NormalTok{vi\_fit}\SpecialCharTok{$}\FunctionTok{summary}\NormalTok{(}\StringTok{"beta"}\NormalTok{) }

\DocumentationTok{\#\# Posterior interval estimates for the probability of labor participation }
\NormalTok{prob\_interval }\OtherTok{\textless{}{-}} \ControlFlowTok{function}\NormalTok{(x, post)\{}
\NormalTok{  lp }\OtherTok{\textless{}{-}}\NormalTok{ post[, }\DecValTok{1}\NormalTok{] }\SpecialCharTok{+}\NormalTok{ x }\SpecialCharTok{*}\NormalTok{ post[, }\DecValTok{2}\NormalTok{]}
  \FunctionTok{quantile}\NormalTok{(}\FunctionTok{exp}\NormalTok{(lp) }\SpecialCharTok{/}\NormalTok{ (}\DecValTok{1} \SpecialCharTok{+} \FunctionTok{exp}\NormalTok{(lp)),}
           \FunctionTok{c}\NormalTok{(.}\DecValTok{05}\NormalTok{, .}\DecValTok{50}\NormalTok{, .}\DecValTok{95}\NormalTok{))}
\NormalTok{\}}

\NormalTok{out }\OtherTok{\textless{}{-}} \FunctionTok{sapply}\NormalTok{(}\FunctionTok{seq}\NormalTok{(}\DecValTok{10}\NormalTok{, }\DecValTok{70}\NormalTok{, }\AttributeTok{by =} \DecValTok{10}\NormalTok{),}
\NormalTok{              prob\_interval, vi\_fit}\SpecialCharTok{$}\FunctionTok{draws}\NormalTok{()[,}\DecValTok{3}\SpecialCharTok{:}\DecValTok{4}\NormalTok{])}

\NormalTok{df\_out }\OtherTok{\textless{}{-}} \FunctionTok{data.frame}\NormalTok{(}\AttributeTok{Income =} \FunctionTok{seq}\NormalTok{(}\DecValTok{10}\NormalTok{, }\DecValTok{70}\NormalTok{, }\AttributeTok{by =} \DecValTok{10}\NormalTok{),}
                     \AttributeTok{Low =}\NormalTok{ out[}\DecValTok{1}\NormalTok{, ],}
                     \AttributeTok{M =}\NormalTok{ out[}\DecValTok{2}\NormalTok{, ],}
                     \AttributeTok{Hi =}\NormalTok{ out[}\DecValTok{3}\NormalTok{,  ])}

\FunctionTok{ggplot}\NormalTok{(df\_out) }\SpecialCharTok{+}
  \FunctionTok{geom\_line}\NormalTok{(}\FunctionTok{aes}\NormalTok{(}\AttributeTok{x =}\NormalTok{ Income, }\AttributeTok{y =}\NormalTok{ M), }\AttributeTok{lwd=}\FloatTok{1.5}\NormalTok{) }\SpecialCharTok{+}
  \FunctionTok{geom\_segment}\NormalTok{(}\FunctionTok{aes}\NormalTok{(}\AttributeTok{x =}\NormalTok{ Income, }\AttributeTok{y =}\NormalTok{ Low,}
                   \AttributeTok{xend =}\NormalTok{ Income, }\AttributeTok{yend =}\NormalTok{ Hi), }\AttributeTok{size =} \DecValTok{2}\NormalTok{) }\SpecialCharTok{+}
  \FunctionTok{ylab}\NormalTok{(}\StringTok{"Prob(Participate)"}\NormalTok{) }\SpecialCharTok{+}
  \FunctionTok{ylim}\NormalTok{(}\DecValTok{0}\NormalTok{, }\DecValTok{1}\NormalTok{) }\SpecialCharTok{+} 
  \FunctionTok{theme}\NormalTok{(}\AttributeTok{text =} \FunctionTok{element\_text}\NormalTok{(}\AttributeTok{size =} \DecValTok{20}\NormalTok{),}
        \AttributeTok{panel.background =} \FunctionTok{element\_rect}\NormalTok{(}\AttributeTok{fill =} \StringTok{"transparent"}\NormalTok{,}
                                        \AttributeTok{color =} \StringTok{"lightgrey"}\NormalTok{),}
        \AttributeTok{panel.grid.major =} \FunctionTok{element\_line}\NormalTok{(}\AttributeTok{colour =} \StringTok{"lightgrey"}\NormalTok{))}

\DocumentationTok{\#\# MCMC fit}
\NormalTok{mcmc\_fit }\OtherTok{\textless{}{-}}\NormalTok{ Logistic\_model\_cmd}\SpecialCharTok{$}\FunctionTok{sample}\NormalTok{(}
  \AttributeTok{data =}\NormalTok{ data, }
  \AttributeTok{seed =} \DecValTok{1}\NormalTok{, }
  \AttributeTok{chains =} \DecValTok{1}\NormalTok{, }
  \AttributeTok{iter\_warmup =} \DecValTok{5000}\NormalTok{,}
  \AttributeTok{iter\_sampling =} \DecValTok{5000}
\NormalTok{)}

\NormalTok{mcmc\_fit}\SpecialCharTok{$}\FunctionTok{summary}\NormalTok{()}

\DocumentationTok{\#\# VI/MCMC comparison}
\NormalTok{df }\OtherTok{\textless{}{-}} \FunctionTok{data.frame}\NormalTok{(}\AttributeTok{Method =} \FunctionTok{c}\NormalTok{(}\FunctionTok{rep}\NormalTok{(}\StringTok{\textquotesingle{}VI\textquotesingle{}}\NormalTok{, }\DecValTok{5000}\NormalTok{), }\FunctionTok{rep}\NormalTok{(}\StringTok{\textquotesingle{}MCMC\textquotesingle{}}\NormalTok{, }\DecValTok{5000}\NormalTok{) ),}
                 \AttributeTok{value =} \FunctionTok{c}\NormalTok{(vi\_fit}\SpecialCharTok{$}\FunctionTok{draws}\NormalTok{(}\StringTok{"alpha"}\NormalTok{),}
\NormalTok{                           mcmc\_fit}\SpecialCharTok{$}\FunctionTok{draws}\NormalTok{(}\StringTok{"alpha"}\NormalTok{)))}

\FunctionTok{ggplot}\NormalTok{(df, }\FunctionTok{aes}\NormalTok{(}\AttributeTok{x=}\NormalTok{value, }\AttributeTok{fill=}\NormalTok{Method)) }\SpecialCharTok{+}
  \FunctionTok{geom\_histogram}\NormalTok{( }\AttributeTok{color=}\StringTok{\textquotesingle{}\#e9ecef\textquotesingle{}}\NormalTok{, }\AttributeTok{alpha=}\FloatTok{0.6}\NormalTok{, }\AttributeTok{position=}\StringTok{\textquotesingle{}identity\textquotesingle{}}\NormalTok{) }\SpecialCharTok{+} 
  \FunctionTok{xlab}\NormalTok{(}\StringTok{"alpha"}\NormalTok{)}

\NormalTok{df }\OtherTok{\textless{}{-}} \FunctionTok{data.frame}\NormalTok{(}\AttributeTok{Method =} \FunctionTok{c}\NormalTok{(}\FunctionTok{rep}\NormalTok{(}\StringTok{\textquotesingle{}VI\textquotesingle{}}\NormalTok{, }\DecValTok{5000}\NormalTok{), }\FunctionTok{rep}\NormalTok{(}\StringTok{\textquotesingle{}MCMC\textquotesingle{}}\NormalTok{, }\DecValTok{5000}\NormalTok{) ),}
                 \AttributeTok{value =} \FunctionTok{c}\NormalTok{(vi\_fit}\SpecialCharTok{$}\FunctionTok{draws}\NormalTok{(}\StringTok{"beta"}\NormalTok{),}
\NormalTok{                           mcmc\_fit}\SpecialCharTok{$}\FunctionTok{draws}\NormalTok{(}\StringTok{"beta"}\NormalTok{)))}

\FunctionTok{ggplot}\NormalTok{(df, }\FunctionTok{aes}\NormalTok{(}\AttributeTok{x=}\NormalTok{value, }\AttributeTok{fill=}\NormalTok{Method)) }\SpecialCharTok{+}
  \FunctionTok{geom\_histogram}\NormalTok{( }\AttributeTok{color=}\StringTok{\textquotesingle{}\#e9ecef\textquotesingle{}}\NormalTok{, }\AttributeTok{alpha=}\FloatTok{0.6}\NormalTok{, }\AttributeTok{position=}\StringTok{\textquotesingle{}identity\textquotesingle{}}\NormalTok{) }\SpecialCharTok{+} 
  \FunctionTok{xlab}\NormalTok{(}\StringTok{"beta"}\NormalTok{)}

\DocumentationTok{\#\# Speed comparison}
\CommentTok{\# VI}
\NormalTok{PSID\_rep }\OtherTok{\textless{}{-}}\NormalTok{ PSID[}\FunctionTok{rep}\NormalTok{(}\FunctionTok{seq\_len}\NormalTok{(}\FunctionTok{nrow}\NormalTok{(PSID)), }\DecValTok{50}\NormalTok{), ]}

\NormalTok{data }\OtherTok{=} \FunctionTok{list}\NormalTok{(}\AttributeTok{N =} \FunctionTok{dim}\NormalTok{(PSID\_rep)[}\DecValTok{1}\NormalTok{],}
            \AttributeTok{x =}\NormalTok{ PSID\_rep}\SpecialCharTok{$}\NormalTok{FamilyIncome,}
            \AttributeTok{y =}\NormalTok{ PSID\_rep}\SpecialCharTok{$}\NormalTok{Participation}
\NormalTok{)}

\NormalTok{vi\_fit }\OtherTok{\textless{}{-}}\NormalTok{ Logistic\_model\_cmd}\SpecialCharTok{$}\FunctionTok{variational}\NormalTok{(}\AttributeTok{data =}\NormalTok{ data,}
                                         \AttributeTok{seed =} \DecValTok{1}\NormalTok{,}
                                         \AttributeTok{output\_samples =} \DecValTok{5000}\NormalTok{,}
                                         \AttributeTok{eval\_elbo =} \DecValTok{1}\NormalTok{,}
                                         \AttributeTok{grad\_samples =} \DecValTok{15}\NormalTok{,}
                                         \AttributeTok{elbo\_samples =} \DecValTok{15}\NormalTok{,}
                                         \AttributeTok{algorithm =} \StringTok{"meanfield"}\NormalTok{,}
                                         \AttributeTok{output\_dir =} \ConstantTok{NULL}\NormalTok{,}
                                         \AttributeTok{iter =} \DecValTok{1000}\NormalTok{,}
                                         \AttributeTok{adapt\_iter =} \DecValTok{20}\NormalTok{,}
                                         \AttributeTok{save\_latent\_dynamics=}\ConstantTok{TRUE}\NormalTok{,}
                                         \AttributeTok{tol\_rel\_obj =} \DecValTok{10}\SpecialCharTok{\^{}{-}}\DecValTok{4}\NormalTok{) }

\CommentTok{\# MCMC}
\NormalTok{mcmc\_fit }\OtherTok{\textless{}{-}}\NormalTok{ Logistic\_model\_cmd}\SpecialCharTok{$}\FunctionTok{sample}\NormalTok{(}
  \AttributeTok{data =}\NormalTok{ data, }
  \AttributeTok{seed =} \DecValTok{1}\NormalTok{, }
  \AttributeTok{chains =} \DecValTok{1}\NormalTok{, }
  \AttributeTok{iter\_warmup =} \DecValTok{5000}\NormalTok{,}
  \AttributeTok{iter\_sampling =} \DecValTok{5000}
\NormalTok{)}
\end{Highlighting}
\end{Shaded}

\hypertarget{lab-on-document-clustering}{%
\section{Lab on Document Clustering}\label{lab-on-document-clustering}}

The goal of this lab is to apply variational inference in a more advanced case study based on Latent Dirichlet Allocation (LDA) and implement the model in \texttt{R} applied to a dataset of documents.
To do so, we consider a collection of 2246 Associated Press newspaper articles to be clustered using the LDA model. The dataset is part of the \texttt{topicmodels R} package. You can load the dataset \texttt{AssociatedPress} with the following R command.

\begin{Shaded}
\begin{Highlighting}[]
\FunctionTok{data}\NormalTok{(}\StringTok{"AssociatedPress"}\NormalTok{, }\AttributeTok{package =} \StringTok{"topicmodels"}\NormalTok{)}
\end{Highlighting}
\end{Shaded}

\hypertarget{application-LDAmodel}{%
\subsection{Overview of the LDA Model and Stan Script}\label{application-LDAmodel}}

The LDA is a mixed-membership clustering model, commonly used for document clustering. LDA models each document to have a mixture of topics, where each word in the document is drawn from a topic based on the mixing proportions. Specifically, the LDA model assumes \(K\) topics for \(M\) documents made up of words drawn from \(V\) distinct words. For document \(m\), a topic distribution \(\boldsymbol{\theta_m}\) is drawn over \(K\) topics from a Dirichlet distribution,
\begin{equation}
\boldsymbol{\theta}_m \sim \textrm{Dirichlet}(\boldsymbol{\alpha}),
\label{eq:theta}
\end{equation}
where \(\sum_{k=1}^{K}\theta_{m, k} = 1\) (\(0 \leq \theta_{m, k} \leq 1\)) and \(\boldsymbol{\alpha}\) is the prior a vector of length \(K\) with positive values.

Each of the \(N_m\) words \(\{w_{m, 1},\dots, w_{m, N_m}\}\) in document \(m\) is then generated independently conditional on \(\boldsymbol{\theta}_m\). To do so, first, the topic \(z_{m, n}\) for word \(w_{m, n}\) in document \(m\) is drawn from
\begin{equation}
z_{m, n} \sim \textrm{categorical}(\boldsymbol{\theta}_m),
\label{eq:z}
\end{equation}
where \(\boldsymbol{\theta}_m\) is the document-specific topic-distribution defined in Equation \eqref{eq:theta}.

Next, the word \(w_{m, n}\) in document \(m\) is drawn from
\begin{equation}
w_{m, n} \sim \textrm{categorical}(\boldsymbol{\phi}_{z[m, n]}),
\label{eq:w}
\end{equation}
which is the word distribution for topic \(z_{m, n}\). Note that z{[}m, n{]} in Equation \eqref{eq:w} refers to \(z_{m, n}\).

Lastly, a Dirichlet prior is given to distributions \(\boldsymbol{\phi}_k\) over words for topic \(k\) as
\begin{equation}
\boldsymbol{\phi}_k \sim \textrm{Dirichlet}(\boldsymbol{\beta}),
\label{eq:phi}
\end{equation}
where \(\boldsymbol{\beta}\) is the prior a vector of length \(V\) (i.e., the total number of words) with positive values. Figure \ref{fig:LDA-model} shows a graphical model representation of LDA.

\begin{figure}

{\centering \includegraphics[width=0.9\linewidth]{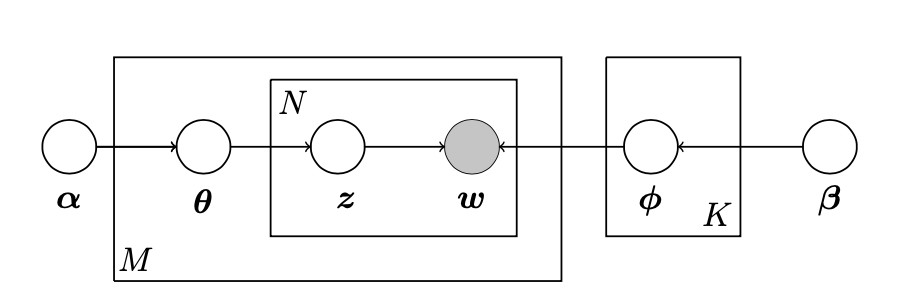} 

}

\caption{Graphical model representation of LDA. The outer box represents the documents. On the left, the inner box represents the topics and words within each document. On the right, the box represents the topics.}\label{fig:LDA-model}
\end{figure}

Below, we include the \texttt{Stan} script for the LDA model provided by \citet{stan_development_team_stan_2012}. Note that \texttt{Stan} supports the calculation of marginal distributions over the continuous parameters by summing out the discrete parameters in mixture models \citep{stan_development_team_stan_2012}. This process corresponds to the \texttt{gamma} parameter in the \texttt{Stan} script below. We refer interested readers to \citet{stan_development_team_stan_2012} for further details.

\begin{Shaded}
\begin{Highlighting}[]
\KeywordTok{data}\NormalTok{ \{}
  \DataTypeTok{int}\NormalTok{\textless{}}\KeywordTok{lower}\NormalTok{=}\DecValTok{2}\NormalTok{\textgreater{} K;               }\CommentTok{// number of topics}
  \DataTypeTok{int}\NormalTok{\textless{}}\KeywordTok{lower}\NormalTok{=}\DecValTok{2}\NormalTok{\textgreater{} V;               }\CommentTok{// number of words}
  \DataTypeTok{int}\NormalTok{\textless{}}\KeywordTok{lower}\NormalTok{=}\DecValTok{1}\NormalTok{\textgreater{} M;               }\CommentTok{// number of docs}
  \DataTypeTok{int}\NormalTok{\textless{}}\KeywordTok{lower}\NormalTok{=}\DecValTok{1}\NormalTok{\textgreater{} N;               }\CommentTok{// total word instances}
  \DataTypeTok{int}\NormalTok{\textless{}}\KeywordTok{lower}\NormalTok{=}\DecValTok{1}\NormalTok{, }\KeywordTok{upper}\NormalTok{=V\textgreater{} w[N];    }\CommentTok{// word n}
  \DataTypeTok{int}\NormalTok{\textless{}}\KeywordTok{lower}\NormalTok{=}\DecValTok{1}\NormalTok{, }\KeywordTok{upper}\NormalTok{=M\textgreater{} doc[N];  }\CommentTok{// doc ID for word n}
  \DataTypeTok{vector}\NormalTok{\textless{}}\KeywordTok{lower}\NormalTok{=}\DecValTok{0}\NormalTok{\textgreater{}[K] alpha;     }\CommentTok{// topic prior vector of length K}
  \DataTypeTok{vector}\NormalTok{\textless{}}\KeywordTok{lower}\NormalTok{=}\DecValTok{0}\NormalTok{\textgreater{}[V] beta;      }\CommentTok{// word prior vector of length V}
\NormalTok{\}}

\KeywordTok{parameters}\NormalTok{ \{}
  \DataTypeTok{simplex}\NormalTok{[K] theta[M];   }\CommentTok{// topic distribution for doc m}
  \DataTypeTok{simplex}\NormalTok{[V] phi[K];     }\CommentTok{// word distribution for topic k}
\NormalTok{\}}

\KeywordTok{model}\NormalTok{ \{}
  \ControlFlowTok{for}\NormalTok{ (m }\ControlFlowTok{in} \DecValTok{1}\NormalTok{:M)}
\NormalTok{    theta[m] \textasciitilde{} dirichlet(alpha);  }
  \ControlFlowTok{for}\NormalTok{ (k }\ControlFlowTok{in} \DecValTok{1}\NormalTok{:K)}
\NormalTok{    phi[k] \textasciitilde{} dirichlet(beta);   }
  \ControlFlowTok{for}\NormalTok{ (n }\ControlFlowTok{in} \DecValTok{1}\NormalTok{:N) \{}
    \DataTypeTok{real}\NormalTok{ gamma[K];}
    \ControlFlowTok{for}\NormalTok{ (k }\ControlFlowTok{in} \DecValTok{1}\NormalTok{:K)}
\NormalTok{      gamma[k] = log(theta[doc[n], k]) + log(phi[k, w[n]]);}
    \KeywordTok{target +=}\NormalTok{ log\_sum\_exp(gamma);  }\CommentTok{// likelihood;}
\NormalTok{  \}  }
\NormalTok{\}}
\end{Highlighting}
\end{Shaded}

\hypertarget{Lab-LDAmodel}{%
\subsection{Variational Inference with the LDA model}\label{Lab-LDAmodel}}

For demonstration purposes, we shall start with a two-topic LDA model (i.e., \(K = 2\)). Before that, we recommend removing the words from \texttt{AssociatedPress} datasets that are rare using the function \texttt{removeSparseTerms()} from the \texttt{tm} package. These words have a minimal effect on the LDA parameter estimation. Nevertheless, they increase the computational cost of variational inference and therefore should be removed using the following R command.

\begin{Shaded}
\begin{Highlighting}[]
\NormalTok{dtm }\OtherTok{\textless{}{-}} \FunctionTok{removeSparseTerms}\NormalTok{(AssociatedPress, }\FloatTok{0.95}\NormalTok{)}
\end{Highlighting}
\end{Shaded}

We are now ready to fit the LDA model using variational inference capabilities of the \texttt{cmdstanr} package. The following code achieves the goal:

\begin{Shaded}
\begin{Highlighting}[]
\NormalTok{LDA\_model\_cmd }\OtherTok{\textless{}{-}} \FunctionTok{cmdstan\_model}\NormalTok{(}\AttributeTok{stan\_file =} \StringTok{"LDA.stan"}\NormalTok{)}

\NormalTok{N\_TOPICS }\OtherTok{\textless{}{-}} \DecValTok{2}
\NormalTok{data }\OtherTok{\textless{}{-}} \FunctionTok{list}\NormalTok{(}\AttributeTok{K =}\NormalTok{ N\_TOPICS,}
             \AttributeTok{V =} \FunctionTok{dim}\NormalTok{(dtm)[}\DecValTok{2}\NormalTok{],}
             \AttributeTok{M =} \FunctionTok{dim}\NormalTok{(dtm)[}\DecValTok{1}\NormalTok{],}
             \AttributeTok{N =} \FunctionTok{sum}\NormalTok{(dtm}\SpecialCharTok{$}\NormalTok{v),}
             \AttributeTok{w =} \FunctionTok{rep}\NormalTok{(dtm}\SpecialCharTok{$}\NormalTok{j,dtm}\SpecialCharTok{$}\NormalTok{v),}
             \AttributeTok{doc =} \FunctionTok{rep}\NormalTok{(dtm}\SpecialCharTok{$}\NormalTok{i,dtm}\SpecialCharTok{$}\NormalTok{v),}
             \CommentTok{\#according to Griffiths and Steyvers(2004) }
             \AttributeTok{alpha =} \FunctionTok{rep}\NormalTok{(}\DecValTok{50}\SpecialCharTok{/}\NormalTok{N\_TOPICS,N\_TOPICS), }
             \AttributeTok{beta =} \FunctionTok{rep}\NormalTok{(}\DecValTok{1}\NormalTok{,}\FunctionTok{dim}\NormalTok{(dtm)[}\DecValTok{2}\NormalTok{]) }
\NormalTok{)}

\NormalTok{vi\_fit }\OtherTok{\textless{}{-}}\NormalTok{ LDA\_model\_cmd}\SpecialCharTok{$}\FunctionTok{variational}\NormalTok{(}\AttributeTok{data =}\NormalTok{ data,}
                                    \AttributeTok{seed =} \DecValTok{1}\NormalTok{,}
                                    \AttributeTok{output\_samples =} \DecValTok{1000}\NormalTok{,}
                                    \AttributeTok{eval\_elbo =} \DecValTok{1}\NormalTok{,}
                                    \AttributeTok{grad\_samples =} \DecValTok{10}\NormalTok{,}
                                    \AttributeTok{elbo\_samples =} \DecValTok{10}\NormalTok{,}
                                    \AttributeTok{algorithm =} \StringTok{"meanfield"}\NormalTok{,}
                                    \AttributeTok{output\_dir =} \ConstantTok{NULL}\NormalTok{,}
                                    \AttributeTok{iter =} \DecValTok{1000}\NormalTok{,}
                                    \AttributeTok{adapt\_iter =} \DecValTok{20}\NormalTok{,}
                                    \AttributeTok{save\_latent\_dynamics=}\ConstantTok{TRUE}\NormalTok{,}
                                    \AttributeTok{tol\_rel\_obj =} \DecValTok{10}\SpecialCharTok{\^{}{-}}\DecValTok{4}\NormalTok{) }
\end{Highlighting}
\end{Shaded}

The ``LDA.stan'' file contains the \texttt{Stan} script for the LDA model provided in Section \ref{application-LDAmodel}. We recommend the usage of the \texttt{R} help to get familiar with the \texttt{variational()} method of the \texttt{cmdstan\_model()} function. The variable \texttt{vi\_fit} contains the results of variational approximation of the LDA parameters. For example, one can obtain the word distributions for the each of the topics with \texttt{vi\_fit\$summary("phi")}.

Finally, to access the ELBO values, use the following:

\begin{Shaded}
\begin{Highlighting}[]
\NormalTok{vi\_diag }\OtherTok{\textless{}{-}}\NormalTok{ utils}\SpecialCharTok{::}\FunctionTok{read.csv}\NormalTok{(vi\_fit}\SpecialCharTok{$}\FunctionTok{latent\_dynamics\_files}\NormalTok{()[}\DecValTok{1}\NormalTok{],}
                           \AttributeTok{comment.char =} \StringTok{"\#"}\NormalTok{)}
\NormalTok{ELBO }\OtherTok{\textless{}{-}} \FunctionTok{data.frame}\NormalTok{(}\AttributeTok{Iteration =}\NormalTok{ vi\_diag[,}\DecValTok{1}\NormalTok{], }\AttributeTok{ELBO =}\NormalTok{ vi\_diag[,}\DecValTok{3}\NormalTok{])}
\end{Highlighting}
\end{Shaded}

\hypertarget{task-1-use-a-graphical-display-to-show-the-10-most-common-words-for-each-of-the-two-topics-and-their-probabilities.}{%
\subsubsection*{Task 1: Use a graphical display to show the 10 most common words for each of the two-topics and their probabilities.}\label{task-1-use-a-graphical-display-to-show-the-10-most-common-words-for-each-of-the-two-topics-and-their-probabilities.}}
\addcontentsline{toc}{subsubsection}{Task 1: Use a graphical display to show the 10 most common words for each of the two-topics and their probabilities.}

\hypertarget{task-2-use-the-function-from-the-package-and-display-the-most-common-words-for-each-of-the-topics-as-a-world-clowd.-what-kinds-of-articles-do-these-topics-represent}{%
\subsubsection*{\texorpdfstring{Task 2: Use the function \texttt{wordcloud()} from the \texttt{wordcloud} package and display the most common words for each of the topics as a world clowd. What kinds of articles do these topics represent?}{Task 2: Use the function  from the  package and display the most common words for each of the topics as a world clowd. What kinds of articles do these topics represent?}}\label{task-2-use-the-function-from-the-package-and-display-the-most-common-words-for-each-of-the-topics-as-a-world-clowd.-what-kinds-of-articles-do-these-topics-represent}}
\addcontentsline{toc}{subsubsection}{Task 2: Use the function \texttt{wordcloud()} from the \texttt{wordcloud} package and display the most common words for each of the topics as a world clowd. What kinds of articles do these topics represent?}

\hypertarget{task-3-fit-a-three-topic-lda-model-display-the-most-common-words-for-each-of-the-topics.-how-do-the-results-differ-from-the-two-topic-lda}{%
\subsubsection*{Task 3: Fit a three-topic LDA model, display the most common words for each of the topics. How do the results differ from the two-topic LDA?}\label{task-3-fit-a-three-topic-lda-model-display-the-most-common-words-for-each-of-the-topics.-how-do-the-results-differ-from-the-two-topic-lda}}
\addcontentsline{toc}{subsubsection}{Task 3: Fit a three-topic LDA model, display the most common words for each of the topics. How do the results differ from the two-topic LDA?}

\hypertarget{task-4-advanced-use-the-three-topic-lda-model-and-diplay-the-topic-prevalence-among-the-2246-associated-press-articles.-that-is-show-what-proportions-of-articles-fall-under-each-topic.}{%
\subsubsection*{Task 4 (Advanced): Use the three-topic LDA model and diplay the topic prevalence among the 2246 Associated Press articles. That is, show what proportions of articles fall under each topic.}\label{task-4-advanced-use-the-three-topic-lda-model-and-diplay-the-topic-prevalence-among-the-2246-associated-press-articles.-that-is-show-what-proportions-of-articles-fall-under-each-topic.}}
\addcontentsline{toc}{subsubsection}{Task 4 (Advanced): Use the three-topic LDA model and diplay the topic prevalence among the 2246 Associated Press articles. That is, show what proportions of articles fall under each topic.}

All necessary R code for fitting the LDA model to the Associated Press sample, including the graphical displays shown in the main text, is included in a separate R script file called LDA\_LAB.R available as a part of the supplementary materials. We also include a printout of the R script below for interested readers.

\begin{Shaded}
\begin{Highlighting}[]
\FunctionTok{library}\NormalTok{(cmdstanr)}

\CommentTok{\# Checking integrity of installation of cmdstanr}
\FunctionTok{check\_cmdstan\_toolchain}\NormalTok{()}
\FunctionTok{install\_cmdstan}\NormalTok{(}\AttributeTok{cores =} \DecValTok{2}\NormalTok{)}
\FunctionTok{cmdstan\_path}\NormalTok{()}
\FunctionTok{cmdstan\_version}\NormalTok{()}

\CommentTok{\# Auxiliary packages}
\FunctionTok{library}\NormalTok{(tm)}
\FunctionTok{library}\NormalTok{(tidyverse)}
\FunctionTok{library}\NormalTok{(tidytext)}
\FunctionTok{library}\NormalTok{(topicmodels)}

\DocumentationTok{\#\# Get data}
\FunctionTok{data}\NormalTok{(}\StringTok{"AssociatedPress"}\NormalTok{, }\AttributeTok{package =} \StringTok{"topicmodels"}\NormalTok{)}

\DocumentationTok{\#\# Removing rare words from the vocabulary}
\NormalTok{dtm }\OtherTok{\textless{}{-}} \FunctionTok{removeSparseTerms}\NormalTok{(AssociatedPress, }\FloatTok{0.95}\NormalTok{)}
\FunctionTok{dim}\NormalTok{(dtm)}

\DocumentationTok{\#\# Input for stan model}
\NormalTok{N\_TOPICS }\OtherTok{\textless{}{-}} \DecValTok{2}

\NormalTok{data }\OtherTok{\textless{}{-}} \FunctionTok{list}\NormalTok{(}\AttributeTok{K =}\NormalTok{ N\_TOPICS,}
             \AttributeTok{V =} \FunctionTok{dim}\NormalTok{(dtm)[}\DecValTok{2}\NormalTok{],}
             \AttributeTok{M =} \FunctionTok{dim}\NormalTok{(dtm)[}\DecValTok{1}\NormalTok{],}
             \AttributeTok{N =} \FunctionTok{sum}\NormalTok{(dtm}\SpecialCharTok{$}\NormalTok{v),}
             \AttributeTok{w =} \FunctionTok{rep}\NormalTok{(dtm}\SpecialCharTok{$}\NormalTok{j,dtm}\SpecialCharTok{$}\NormalTok{v),}
             \AttributeTok{doc =} \FunctionTok{rep}\NormalTok{(dtm}\SpecialCharTok{$}\NormalTok{i,dtm}\SpecialCharTok{$}\NormalTok{v),}
             \CommentTok{\#according to Griffiths and Steyvers(2004) }
             \AttributeTok{alpha =} \FunctionTok{rep}\NormalTok{(}\DecValTok{50}\SpecialCharTok{/}\NormalTok{N\_TOPICS,N\_TOPICS), }
             \AttributeTok{beta =} \FunctionTok{rep}\NormalTok{(}\DecValTok{1}\NormalTok{,}\FunctionTok{dim}\NormalTok{(dtm)[}\DecValTok{2}\NormalTok{])  }
\NormalTok{)}

\DocumentationTok{\#\#\# VB fit}
\NormalTok{LDA\_model\_cmd }\OtherTok{\textless{}{-}} \FunctionTok{cmdstan\_model}\NormalTok{(}\AttributeTok{stan\_file =} \StringTok{"LDA.stan"}\NormalTok{)}
\NormalTok{LDA\_model\_cmd}\SpecialCharTok{$}\FunctionTok{print}\NormalTok{()  }

\NormalTok{vb\_fit }\OtherTok{\textless{}{-}}\NormalTok{ LDA\_model\_cmd}\SpecialCharTok{$}\FunctionTok{variational}\NormalTok{(}\AttributeTok{data =}\NormalTok{ data,}
                                    \AttributeTok{seed =} \DecValTok{1}\NormalTok{,}
                                    \AttributeTok{output\_samples =} \DecValTok{1000}\NormalTok{,}
                                    \AttributeTok{eval\_elbo =} \DecValTok{1}\NormalTok{,}
                                    \AttributeTok{grad\_samples =} \DecValTok{10}\NormalTok{,}
                                    \AttributeTok{elbo\_samples =} \DecValTok{10}\NormalTok{,}
                                    \AttributeTok{algorithm =} \StringTok{"meanfield"}\NormalTok{,}
                                    \AttributeTok{output\_dir =} \ConstantTok{NULL}\NormalTok{,}
                                    \AttributeTok{iter =} \DecValTok{1000}\NormalTok{,}
                                    \AttributeTok{adapt\_iter =} \DecValTok{20}\NormalTok{,}
                                    \AttributeTok{save\_latent\_dynamics=}\ConstantTok{TRUE}\NormalTok{,}
                                    \AttributeTok{tol\_rel\_obj =} \DecValTok{10}\SpecialCharTok{\^{}{-}}\DecValTok{4}\NormalTok{) }

\CommentTok{\# Plotting ELBO}
\NormalTok{vb\_diag }\OtherTok{\textless{}{-}}\NormalTok{ utils}\SpecialCharTok{::}\FunctionTok{read.csv}\NormalTok{(vb\_fit}\SpecialCharTok{$}\FunctionTok{latent\_dynamics\_files}\NormalTok{()[}\DecValTok{1}\NormalTok{], }
                           \AttributeTok{comment.char =} \StringTok{"\#"}\NormalTok{)}
\NormalTok{ELBO }\OtherTok{\textless{}{-}} \FunctionTok{data.frame}\NormalTok{(}\AttributeTok{Iteration =}\NormalTok{ vb\_diag[,}\DecValTok{1}\NormalTok{], }
                  \AttributeTok{ELBO =}\NormalTok{ vb\_diag[,}\DecValTok{3}\NormalTok{])}

\FunctionTok{ggplot}\NormalTok{(}\AttributeTok{data =}\NormalTok{ ELBO, }\FunctionTok{aes}\NormalTok{(}\AttributeTok{x =}\NormalTok{ Iteration, }\AttributeTok{y =}\NormalTok{ ELBO)) }\SpecialCharTok{+} \FunctionTok{geom\_line}\NormalTok{(}\AttributeTok{lwd=}\FloatTok{1.5}\NormalTok{) }\SpecialCharTok{+} 
  \FunctionTok{theme}\NormalTok{(}\AttributeTok{text =} \FunctionTok{element\_text}\NormalTok{(}\AttributeTok{size =} \DecValTok{20}\NormalTok{), }
        \AttributeTok{panel.background =} \FunctionTok{element\_rect}\NormalTok{(}\AttributeTok{fill =} \StringTok{"transparent"}\NormalTok{, }
                                        \AttributeTok{color =} \StringTok{"lightgrey"}\NormalTok{), }
        \AttributeTok{panel.grid.major =} \FunctionTok{element\_line}\NormalTok{(}\AttributeTok{colour =} \StringTok{"lightgrey"}\NormalTok{)) }\SpecialCharTok{+}
  \FunctionTok{xlim}\NormalTok{(}\DecValTok{0}\NormalTok{,}\DecValTok{110}\NormalTok{)}

\DocumentationTok{\#\# Accessing parameters}
\NormalTok{vb\_fit}\SpecialCharTok{$}\FunctionTok{summary}\NormalTok{(}\StringTok{"theta"}\NormalTok{) }\CommentTok{\# dim: M{-}by{-}K}
\NormalTok{vb\_fit}\SpecialCharTok{$}\FunctionTok{summary}\NormalTok{(}\StringTok{"phi"}\NormalTok{) }\CommentTok{\# dim: K{-}by{-}V}

\DocumentationTok{\#\# Word distribution per topic}
\NormalTok{V }\OtherTok{\textless{}{-}} \FunctionTok{dim}\NormalTok{(dtm)[}\DecValTok{2}\NormalTok{]}
\NormalTok{odd\_rows }\OtherTok{\textless{}{-}} \FunctionTok{rep}\NormalTok{(}\FunctionTok{c}\NormalTok{(}\DecValTok{1}\NormalTok{,}\DecValTok{0}\NormalTok{), }\AttributeTok{times =}\NormalTok{ V)}
\NormalTok{Topic1 }\OtherTok{\textless{}{-}}\NormalTok{ vb\_fit}\SpecialCharTok{$}\FunctionTok{summary}\NormalTok{(}\StringTok{"phi"}\NormalTok{)[odd\_rows }\SpecialCharTok{==} \DecValTok{1}\NormalTok{,]}
\NormalTok{Topic2 }\OtherTok{\textless{}{-}}\NormalTok{ vb\_fit}\SpecialCharTok{$}\FunctionTok{summary}\NormalTok{(}\StringTok{"phi"}\NormalTok{)[odd\_rows }\SpecialCharTok{==} \DecValTok{0}\NormalTok{,]}

\NormalTok{word\_probs }\OtherTok{\textless{}{-}} \FunctionTok{data.frame}\NormalTok{(}\AttributeTok{Topic =} \FunctionTok{c}\NormalTok{(}\FunctionTok{rep}\NormalTok{(}\StringTok{"Topic 1"}\NormalTok{, V), }
                                  \FunctionTok{rep}\NormalTok{(}\StringTok{"Topic 2"}\NormalTok{, V)),}
                        \AttributeTok{Word =} \FunctionTok{rep}\NormalTok{(dtm}\SpecialCharTok{$}\NormalTok{dimnames}\SpecialCharTok{$}\NormalTok{Terms,N\_TOPICS),}
                        \AttributeTok{Probability =} \FunctionTok{c}\NormalTok{(Topic1}\SpecialCharTok{$}\NormalTok{mean, Topic2}\SpecialCharTok{$}\NormalTok{mean))}

\CommentTok{\# Selecting top 10 words per topic}
\NormalTok{top\_words }\OtherTok{\textless{}{-}}\NormalTok{ word\_probs }\SpecialCharTok{\%\textgreater{}\%} \FunctionTok{group\_by}\NormalTok{(Topic) }\SpecialCharTok{\%\textgreater{}\%} \FunctionTok{top\_n}\NormalTok{(}\DecValTok{10}\NormalTok{) }\SpecialCharTok{\%\textgreater{}\%} 
  \FunctionTok{ungroup}\NormalTok{() }\SpecialCharTok{\%\textgreater{}\%} \FunctionTok{arrange}\NormalTok{(Topic, }\SpecialCharTok{{-}}\NormalTok{Probability)}

\NormalTok{top\_words }\SpecialCharTok{\%\textgreater{}\%}
  \FunctionTok{mutate}\NormalTok{(}\AttributeTok{Word =} \FunctionTok{reorder\_within}\NormalTok{(Word, Probability, Topic)) }\SpecialCharTok{\%\textgreater{}\%}
  \FunctionTok{ggplot}\NormalTok{(}\FunctionTok{aes}\NormalTok{(Probability, Word, }\AttributeTok{fill =} \FunctionTok{factor}\NormalTok{(Topic))) }\SpecialCharTok{+}
  \FunctionTok{geom\_col}\NormalTok{(}\AttributeTok{show.legend =} \ConstantTok{FALSE}\NormalTok{) }\SpecialCharTok{+}
  \FunctionTok{facet\_wrap}\NormalTok{(}\SpecialCharTok{\textasciitilde{}}\NormalTok{ Topic, }\AttributeTok{scales =} \StringTok{"free"}\NormalTok{) }\SpecialCharTok{+}
  \FunctionTok{scale\_y\_reordered}\NormalTok{() }\SpecialCharTok{+} \FunctionTok{theme}\NormalTok{(}\AttributeTok{text =} \FunctionTok{element\_text}\NormalTok{(}\AttributeTok{size =} \DecValTok{15}\NormalTok{)) }\SpecialCharTok{+} \FunctionTok{xlim}\NormalTok{(}\DecValTok{0}\NormalTok{,}\FloatTok{0.025}\NormalTok{) }\SpecialCharTok{+}
  \FunctionTok{xlab}\NormalTok{(}\StringTok{"Word distributions ( \textbackslash{}u03d5 )"}\NormalTok{)}

\CommentTok{\# Word Cloud display}
\CommentTok{\#install.packages("wordcloud")}
\FunctionTok{library}\NormalTok{(wordcloud)}

\NormalTok{top\_words }\OtherTok{\textless{}{-}}\NormalTok{ word\_probs }\SpecialCharTok{\%\textgreater{}\%} \FunctionTok{group\_by}\NormalTok{(Topic) }\SpecialCharTok{\%\textgreater{}\%} \FunctionTok{top\_n}\NormalTok{(}\DecValTok{20}\NormalTok{) }\SpecialCharTok{\%\textgreater{}\%} 
  \FunctionTok{ungroup}\NormalTok{() }\SpecialCharTok{\%\textgreater{}\%} \FunctionTok{arrange}\NormalTok{(Topic, }\SpecialCharTok{{-}}\NormalTok{Probability)}

\NormalTok{mycolors }\OtherTok{\textless{}{-}} \FunctionTok{brewer.pal}\NormalTok{(}\DecValTok{8}\NormalTok{, }\StringTok{"Dark2"}\NormalTok{)}
\FunctionTok{wordcloud}\NormalTok{(top\_words }\SpecialCharTok{\%\textgreater{}\%} \FunctionTok{filter}\NormalTok{(Topic }\SpecialCharTok{==} \StringTok{"Topic 1"}\NormalTok{) }\SpecialCharTok{\%\textgreater{}\%}\NormalTok{ .}\SpecialCharTok{$}\NormalTok{Word ,}
\NormalTok{          top\_words }\SpecialCharTok{\%\textgreater{}\%} \FunctionTok{filter}\NormalTok{(Topic }\SpecialCharTok{==} \StringTok{"Topic 1"}\NormalTok{) }\SpecialCharTok{\%\textgreater{}\%}\NormalTok{ .}\SpecialCharTok{$}\NormalTok{Probability,}
          \AttributeTok{random.order =} \ConstantTok{FALSE}\NormalTok{,}
          \AttributeTok{color =}\NormalTok{ mycolors)}

\NormalTok{mycolors }\OtherTok{\textless{}{-}} \FunctionTok{brewer.pal}\NormalTok{(}\DecValTok{8}\NormalTok{, }\StringTok{"Dark2"}\NormalTok{)}
\FunctionTok{wordcloud}\NormalTok{(top\_words }\SpecialCharTok{\%\textgreater{}\%} \FunctionTok{filter}\NormalTok{(Topic }\SpecialCharTok{==} \StringTok{"Topic 2"}\NormalTok{) }\SpecialCharTok{\%\textgreater{}\%}\NormalTok{ .}\SpecialCharTok{$}\NormalTok{Word ,}
\NormalTok{          top\_words }\SpecialCharTok{\%\textgreater{}\%} \FunctionTok{filter}\NormalTok{(Topic }\SpecialCharTok{==} \StringTok{"Topic 2"}\NormalTok{) }\SpecialCharTok{\%\textgreater{}\%}\NormalTok{ .}\SpecialCharTok{$}\NormalTok{Probability,}
          \AttributeTok{random.order =} \ConstantTok{FALSE}\NormalTok{,}
          \AttributeTok{color =}\NormalTok{ mycolors)}
\end{Highlighting}
\end{Shaded}

\end{document}